\documentstyle[prd,aps,epsf,eqsecnum]{revtex}
\begin{document}
\def\d{{\mathrm{d}}}
\def\nbox#1#2{\vcenter{\hrule \hbox{\vrule height#2in
\kern#1in \vrule} \hrule}}
\def\sq{\,\raise.5pt\hbox{$\nbox{.09}{.09}$}\,}
\def\sqb{\,\raise.5pt\hbox{$\overline{\nbox{.09}{.09}}$}\,}    
\newcommand{\tab}{$\langle T_{ab} \rangle$}

\title
{Linear Response, Validity of Semi-Classical Gravity, \\ and the
Stability of Flat Space}

\author
{
Paul R. Anderson$^{1}$
\thanks{electronic address: anderson@wfu.edu},
Carmen Molina-Par\'{\i}s$^{2,3}$
\thanks{electronic address: molina@maths.warwick.ac.uk},
and
Emil Mottola$^{2}$
\thanks{electronic address: emil@lanl.gov}
}

\address{
$^{1}$
Department of Physics,
Wake Forest University, Winston-Salem,
North Carolina, 27109\\
$^{2}$
T-8, Theoretical Division, Los Alamos
National Laboratory, Los Alamos, New Mexico, 87545
\\
$^{3}$
Mathematics Institute, University of Warwick,
Coventry CV4 7AL, UK}

\date{LA-UR-02-5809}
\maketitle

\abstract 

{A quantitative test for the validity of the semi-classical
approximation in gravity is given. The criterion proposed is that
solutions to the semi-classical Einstein equations should be stable to
linearized perturbations, in the sense that no gauge invariant
perturbation should become unbounded in time. A self-consistent linear
response analysis of these perturbations, based upon an invariant
effective action principle, necessarily involves metric fluctuations
about the mean semi-classical geometry, and brings in the two-point
correlation function of the quantum energy-momentum tensor in a
natural way. This linear response equation contains no state dependent
divergences and requires no new renormalization counterterms beyond
those required in the leading order semi-classical approximation.  The
general linear response criterion is applied to the specific example
of a scalar field with arbitrary mass and curvature coupling in the
vacuum state of Minkowski spacetime. The spectral representation of
the vacuum polarization function is computed in $n$ dimensional
Minkowski spacetime, and used to show that the flat space solution to
the semi-classical Einstein equations for $n=4$ is stable to all
perturbations on distance scales much larger than the Planck length.}

\section{Introduction}
\label{sec:1}

There are many well known difficulties that arise when attempting to
combine quantum field theory and general relativity into a full
quantum theory of gravity. Almost certainly, a consistent quantum
theory at the Planck scale requires a fundamentally different set of
principles from those of classical general relativity, in which even
the concept of spacetime itself is likely to be radically altered.
Yet, over a very wide range of distance scales, from that of the
electroweak interactions ($10^{-16}$ cm) to cosmology ($10^{27}$ cm),
the basic framework of a spacetime metric theory obeying general
coordinate invariance is assumed to be valid, and receives
phenomenological support both from the successes of flat space quantum
field theory at the lower end of this distance scale, and classical
general relativity at its upper end. Hence, whatever the full quantum
theory of gravity entails, it should reduce to an effective low energy
field theory on this very broad range of some $43$ orders of magnitude
of distance~\cite{hu,donoghue}.

To the extent that quantum effects are relevant at all in
gravitational phenomena within this range of scales, one would expect
to be able to apply {\it semi}-classical techniques to the low energy
effective theory of gravity. In the semi-classical approximation to
gravity the spacetime metric $g_{ab}$ is treated as a classical
$c$-number field and its quantum fluctuations are neglected, although
quantum fluctuations of the other fields are taken into account. The
semi-classical approach has been discussed and studied for some time
now, and a considerable body of results has been obtained
\cite{birrell-davies}.  Yet a definitive answer to the question of
what is the limit of validity of this approach, and under what physical
conditions it must break down has remained somewhat unclear.

It is our purpose in this paper to propose a well-defined {\it
quantitative} criterion for the validity of the semi-classical
approximation to gravity, within the semi-classical formalism itself,
namely that solutions to the semi-classical Einstein equations should
be stable against linearized perturbations of the geometry. This
criterion may be formulated by means of a linear response
analysis ~\cite{FW,leb,mot-fd}.

It is important to distinguish what we mean in this paper by the
semi-classical approximation from the ordinary loop expansion, which
is sometimes also called semi-classical. In the ordinary loop
expansion of the effective action, $\hbar$ is the formal (loop
expansion) parameter.  As a result both the matter and gravitational
quantum fluctuations are treated on exactly the same footing, and the
back-reaction of these fluctuations on the metric (being first order
in the expansion parameter $\hbar$) are neglected. If one does attempt
to include such effects in some modified loop expansion, the technical
issues involved in defining a one-loop effective action for gravitons
that respects both linearized gauge and background field coordinate
invariance must be faced. These are difficult enough to have impeded
progress in the standard loop expansion in
gravity~\cite{dewitt,dw-mp}. An unambiguous definition of the
corresponding conserved and gauge invariant energy-momentum tensor for
gravitons in an arbitrary curved spacetime has not yet been
given~\cite{brandenberger}. Apart from such technical difficulties it
should be clear that a simple loop expansion is ill-suited physically
to many applications that have been and are likely to be of interest
in semi-classical gravity, such as particle creation in the early
universe, or black hole radiance, where the quantum effects of matter
significantly affect the background geometry after some period of
time, but (it is usually assumed) the quantum fluctuations of the
geometry itself can be neglected.  Whenever quantum effects of matter
are expected significantly to affect the classical geometry, the
standard loop expansion, which treats these effects as order $\hbar$
and small, must certainly break down.

The semi-classical approximation to gravity we discuss in this paper
treats the matter fields as quantum but the spacetime metric as
classical, and allows for the consistent back-reaction of the quantum
matter on the classical geometry. This asymmetric treatment can be
justified formally by replicating the number of matter fields $N$
times and taking the large $N$ limit of the quantum effective action
for the matter fields in an arbitrary background metric
$g_{ab}$~\cite{tomboulis}. Then, the semi-classical equations for the
metric are derived by varying the effective action, with local
gravitational terms included. Since no assumption of the weakness or
perturbative nature of the metric is assumed, the large $N$ expansion
is able to address problems in which gravitational effects on the
matter are strong, and the matter fields can have a large cumulative
effect on the classical geometry in turn. The absence of quantum
gravitational effects in the lowest order large $N$ approximation also
means that the technical obstacles arising from the quantum
fluctuations of the geometry are avoided. General coordinate
invariance is assured, provided only that the matter effective action
is regularized and renormalized in a manner which respects that
invariance~\cite{dewitt}. In that case the quantum expectation value
of the matter energy-momentum tensor $\langle T_{ab}\rangle$ is
necessarily conserved.

Assuming that the classical energy-momentum tensor for the matter
field(s) vanishes (an assumption that may be easily relaxed if
necessary), the unrenormalized semi-classical back-reaction equations
take the form~\cite{mtw}
\begin{equation}  
G_{ab} + \Lambda g_{ab} = 8\pi G_{_N} \langle T_{ab}\rangle
\; .  
\label{eq:scE} 
\end{equation} 
Here $G_{ab}$ is the Einstein tensor, $\Lambda$ is the cosmological
constant (which may be taken to be zero in some applications),
$G_{_N}$ is Newton's constant, and $\langle T_{ab}\rangle$ is the
expectation value of the energy-momentum tensor operator of the
quantized matter field(s).  Among the technical issues that must be
confronted is the renormalization of the expectation value of
$T_{ab}$, a quartically divergent composite operator in $n=4$
spacetime dimensions. The renormalization of its expectation value
requires the introduction of fourth order counterterms in the
effective action, that modify the geometric terms on the left hand
side of eq.~(\ref{eq:scE})~\cite{birrell-davies}.  

Once a renormalized semi-classical theory has been defined, one
possible route to investigating its validity is to compare
calculations in a theory of quantum gravity with similar
semi-classical calculations. Since a well-defined, full quantum theory
is lacking, this has been done only in some simplified models of
quantum gravity. Large quantum gravity effects were found in
three-dimensional models by Ashtekar~\cite{ashtekar} and
Beetle~\cite{beetle}. In four dimensions, Ford has considered the case
of graviton production in a linearized theory of quantum gravity on a
flat space background, and compared the results with the production of
gravitational waves in semi-classical gravity~\cite{ford}.  He found
that they were comparable when the renormalized energy-momentum
(connected) correlation function,
\begin{equation} 
\langle T_{ab}(x)
T_{cd}(x') \rangle_{\rm con} \equiv \langle T_{ab}(x) T_{cd}(x') \rangle
- \langle T_{ab}(x) \rangle \langle T_{cd}(x') \rangle
\label{eq:TT}
\end{equation}
satisfied the condition
\begin{equation} 
\langle T_{ab}(x) T_{cd}(x') \rangle_{\rm con} \ll 
\langle
T_{ab}(x)  \rangle \langle T_{cd}(x')\rangle 
\; . 
\label{eq:ford} 
\end{equation}

The limits of validity of the semi-classical approximation have also
been studied without making reference to a specific model of quantum
gravity. Kuo and Ford~\cite{kuo-ford} proposed that a measure of how
strongly the semi-classical approximation is violated can be given by
how large the quantity,
\begin{equation} 
\Delta_{abcd} (x,x') \equiv
\left|\frac{\langle T_{ab}(x) T_{cd}(x') \rangle_{\rm con}} {\langle
T_{ab}(x) T_{cd}(x')\rangle}\right|
\label{eq:k-f}
\end{equation}
is, where it is assumed that the expectation values in this expression
are suitably renormalized.  It is important to note that eq.~(\ref{eq:k-f}) 
is coordinate dependent, since both the numerator and denominator are
{\it tensor} quantities. The situation is complicated further by the
regularization and renormalization issues that arise in defining the
quantities appearing in this expression. Using normal ordering, 
Kuo and Ford~\cite{kuo-ford} computed the quantity
\begin{equation} 
\Delta (x) \equiv \left|\frac{\langle T_{00}(x)
T_{00}(x) \rangle_{\rm con}} {\langle T_{00}(x)
T_{00}(x)\rangle}\right| \label{eq:variance}
\end{equation}
for a free scalar field in flat space for several states including the
Casimir vacuum.  They found that it vanishes in a coherent state,
whereas in many other cases, including the Casimir vacuum, it is of
order unity.

Wu and Ford~\cite{wu-ford-1} computed the radial flux component of
eq.~(\ref{eq:k-f}), in the cases of a moving mirror in 2-dimensions
and an evaporating black hole far from the event horizon in both $2$
and $4$ dimensions. They found that it was of order unity over time
scales comparable to the black hole mass, but that it averages to zero
over much larger times. In a normal ordering prescription they found
state dependent divergent terms.  They also showed that in the simple
case of radiation exerting a force on a mirror, the quantum
fluctuations in the radiation pressure are due to a state dependent
cross term in the energy-momentum tensor correlation
function~\cite{wu-ford-2,wu-ford-3}.
 
Phillips and Hu~\cite{phillips-hu-1} used zeta function regularization
to compute $\Delta (x)$ with the denominator replaced by the quantity
$\langle T_{00}(x)\rangle^2$, for a free scalar field in some curved
spacetimes having Euclidean sections. They also computed $\Delta (x)$
for a scalar field in flat space in the Minkowski vacuum state, using
both point splitting and a smearing operator to remove the
divergences~\cite{phillips-hu-2}.  For the flat space calculation they
found that $\Delta (x)$ depends on the direction the points are split,
but that it is of order unity regardless of how the points are split.
They used their results to criticize the Kuo-Ford conjecture and to
suggest that the criteria for the validity of the semi-classical
approximation should depend on the scale at which the system is being
probed.
 
Wu and Ford~\cite{wu-ford-3} addressed the Kuo-Ford conjecture and the
above mentioned criticism of it by Phillips and Hu.  They stated that
the conjecture is incomplete because it does not address the effect of
divergent state dependent terms. They suggested that any criterion for
the validity of the semi-classical approximation should be a non-local
one that involves integrals over the world lines of test
particles. They also argued that the question of whether the
semi-classical approximation is valid depends on the specifics of a
given situation, including the scales being probed and the choice of
initial quantum state.

Although it is somewhat unclear what the dimensionless small parameter
is that controls the inequality~(\ref{eq:ford}), Ford's initial work
and these subsequent discussions draw attention to the importance of
the higher point correlation functions of the energy-momentum tensor.
It is quite clear, at least in qualitative terms, that if the higher
point connected correlation functions of $T_{ab}$ are large (in an
appropriate sense to be determined), it cannot be correct to neglect
them completely, as the semi-classical equations~(\ref{eq:scE})
certainly do.

Another context in which the quantity $\langle T_{ab}(x) T_{cd}(x')
\rangle_{\rm con}$ plays a role is stochastic semi-classical gravity
~\cite{hu,calzetta-hu,hu-matacz,hu-sinha,martin-verdaguer-1,martin-verdaguer-2}.
In this case the probability distribution function for the quantum
noise is obtained from the symmetric part of this correlation
function~\cite{calzetta-hu}. The dissipation kernel has been shown
to be related to the antisymmetric part of this correlation
function~\cite{campos-hu-1}. Stochastic semi-classical gravity is an
interesting attempt to go beyond the semi-classical approximation. However, for
the purposes of the present work, we do not make any stochastic assumptions and
investigate the validity of the semi-classical equations within the large $N$
approximation itself.

The energy-momentum correlation function $\langle T_{ab}(x) T_{cd}(x')
\rangle_{\rm con}$ has been directly computed for a scalar field in a
two dimensional spacetime with a moving boundary~\cite{carlitz-willey}, for
scalar fields and the Maxwell field in Minkowski
spacetime~\cite{jaekel-reynaud,martin-verdaguer-3}, and for a massless
minimally coupled scalar field in de Sitter spacetime, in the case that the
points are spacelike separated and geodesically
connected~\cite{roura-verdaguer}. It has also been computed  indirectly through
the non-local kernel appearing in the deviation of  $\langle T_{ab}\rangle$
from flat space~\cite{capper-duff-halpern,horowitz,jordan-1}, from a
Robertson-Walker spacetime~\cite{hartle-hu}, and from a general conformally
flat spacetime~\cite{horowitz-wald}.   The noise and dissipation kernels in
stochastic semi-classical gravity are related to the energy-momentum tensor
correlation function~\cite{campos-hu-1}.  These quantities have been computed
exactly or approximately for scalar fields of various types in several
situations including Minkowski
spacetime~\cite{campos-verdaguer-1,campos-verdaguer-2}, hot flat
space~\cite{campos-hu-1}, the far field limit of a black hole in equilibrium
with a thermal field~\cite{campos-hu-2}, Robertson-Walker
spacetimes~\cite{calzetta-hu,hu-matacz,calzetta-campos-verdaguer,calzetta-verdaguer}, Bianchi type I spacetimes~\cite{hu-sinha}, and a weakly curved spacetime using a covariant expansion in powers of the curvature~\cite{lombardo-mazzitelli}.

Although technical problems such as renormalization and coordinate
invariance complicate matters, this body of previous work suggests
that the correlation function $\langle T_{ab}(x) T_{cd}(x')
\rangle_{\rm con}$ should play an important role in determining the
validity of the semi-classical approximation.  However, the proper
context for incorporating and making use of the information contained
in this correlation function in a well-defined ({\it i.e.,} finite),
quantitative framework, that respects general coordinate invariance
has remained somewhat unclear.

The criterion we propose in this paper, that solutions to the
semi-classical Einstein equations should be stable against linearized
perturbations of the geometry, provides just such a framework. According to
standard linear response theory~\cite{FW,leb,mar-sch}, the linearized equations
for the perturbed metric depend on the retarded two-point correlation function
of the energy-momentum tensor evaluated in the semi-classical background metric
$g_{ab}$~\cite{mot-fd}.  In this case, the correlation function can be computed
using the Closed Time Path (CTP) effective action~\cite{CTP}. The result is a
retarded correlation function that involves the commutator of two
energy-momentum tensor operators. Hence the perturbations are manifestly
causal. Moreover, the UV divergences found in the unrenormalized linear response
equations are exactly those required to renormalize the semi-classical
theory itself.  This ensures that no state-dependent divergences
occur. Finally, gauge transformations of the linearized metric
fluctuations, $h_{ab}$, are easily handled within the linear response
framework, so that ambiguities related to quantities such
as~(\ref{eq:k-f}) do not arise. Thus, standard linear response
theory provides a well-defined framework in which to test the validity of the
semi-classical approximation to gravity, which directly involves $\langle
T_{ab}(x) T_{cd}(x')\rangle$ and its renormalization, in a manner that is in
complete accordance with the physical principles of general covariance and
causality.

Since this criterion for the validity of the semi-classical
approximation lies strictly within the context of that approximation itself,
one avoids problems such as gauge invariance of the energy-momentum
tensor for gravitons, that inevitably appear if one tries to go beyond
the semi-classical approximation and include quantum effects due to
the gravitational field. Although these effects certainly are not
contained in the semi-classical Einstein eqs.~(\ref{eq:scE}), it is
possible to study the properties of linearized gravitational
fluctuations about the self-consistent solution of eq.~(\ref{eq:scE}),
simply by taking one higher variation of the effective action that
leads to that equation.  This second variation involves the two-point
correlation function~(\ref{eq:TT}), evaluated in the self-consistent
background geometry.

To understand qualitatively the role of the two-point correlation
function in the validity of the semi-classical approximation, it is
helpful to consider the physical analogy between semi-classical
gravity and semi-classical electromagnetism.  The connected
correlation function~(\ref{eq:TT}) measures the gravitational vacuum
polarization, which contributes to the proper self-energy of the
linearized graviton fluctuations around the background metric, just as
the current two-point correlation function, $\langle j^a(x)
j^b(x')\rangle_{\rm con}$, measures the electromagnetic vacuum
polarization which contributes to the proper self-energy of the
photon~\cite{photon,kal-leh}. Hence, if these polarization effects are
significant, the semi-classical approximation has certainly broken
down, at least in the form specified by eq.~(\ref{eq:scE}), where all
fluctuations of the metric have been ignored. In quantum
electromagnetism (QED) we know exactly how to take these fluctuation
effects into account, namely by scattering and interaction Feynman
diagrams involving the photon propagator. These processes are
important not only in scattering between a few particles at high
energies, but also in low energy processes in hot or dense
plasmas~\cite{leb}.  Analogous statements should be applicable to
gravity. Thus, if the linear response validity criterion is not satisfied,
there will be no avoiding the technical difficulties and physical consequences
of treating the fluctuations of the gravitational field itself, even if we seek
to understand only the {\it infrared} behavior of a semi-classical
approximation to the effective theory of gravity, far below the Planck energy
scale.

As a particular illustration of the validity criterion, we apply it to
the example of a scalar field with arbitrary mass and curvature
coupling in the vacuum state of Minkowski spacetime. We express the
retarded correlation function of the linear response analysis in flat
space in terms of a K\"all\'en-Lehmann spectral
representation~\cite{kal-leh}. The positivity of the spectral
representation is sufficient to demonstrate that there are no unstable
modes of the linearized semi-classical equations around flat space at
distance scales far larger than the Planck scale, and hence, that flat
spacetime is completely infrared stable in semi-classical gravity.
The semi-classical stability of Minkowski spacetime has been
investigated previously by several
authors~\cite{horowitz,jordan-1,suen}, and instabilities have
been found which involve strictly Planck scale variations of the metric
fluctuations in space and/or time, which arise from the terms fourth order in
derivatives of the metric that are needed to renormalize eqs.~(\ref{eq:scE}).
Their existence clearly precludes the validity of the semi-classical large $N$
approximation at Planck time or distance scales. Prescriptions for explicitly
reducing the order of the equations, which eliminates these Planck scale
solutions have been proposed~\cite{simon,parker-simon} and 
discussed in some detail~\cite{flanagan-wald}.  Whether or not these
prescriptions are accepted in the general case, it is quite clear {\it a
priori} that the semi-classical approximation~(\ref{eq:scE}) can be viewed at
best only as the low energy effective field theory limit of a more complete
quantum theory~\cite{hu,donoghue}, and that no reliable results can be
obtained from this approximation in the Planckian regime. However, the
flat space example treated in some detail in this paper shows
explicitly that the semi-classical approximation does give
mathematically meaningful and physically sensible results, when
properly restricted to its range of validity at space and distance
scales very much larger than the Planck scale.

The organization of the paper is as follows. In the next Section the
properties of the large $N$ semi-classical approximation in gravity
and its renormalization within the covariant effective action
framework are reviewed. In Section III the linear response theory for
the semi-classical back-reaction equations is described.  The form of
the two-point correlation function for the energy-momentum tensor that
appears in the linear response equations is given, and its properties
and renormalization are discussed. Then our proposal for a necessary
condition for the validity of the semi-classical approximation is
presented. In Section IV the use of our criterion is illustrated for
the case of a scalar field with arbitrary mass and curvature coupling
in the vacuum state in Minkowski spacetime. The linear response
analysis implies that flat spacetime is stable under small
fluctuations at large wavelengths. Our results are discussed further
in Section V. Some additional applications of our criterion to the
study of quantum effects in cosmological and black hole spacetimes are
suggested. There are two Appendices. The first deals with the general
decomposition of tensors and polarization operators in Minkowski
spacetime. The second contains the technical details of the
computation of the retarded correlation response function for a scalar
field in Minkowski spacetime.

\section{Semi-Classical Gravity and Renormalization}
\label{sec:2}

The most direct route to the semi-classical equations~(\ref{eq:scE})
is via the effective action method in the large $N$ limit. We consider
the specific example of $N$ non-interacting scalar fields.
Generalizations to interacting fields and fields of other spin are
straightforward, but as they are not required to expose the main
elements of the stability criterion, we treat only this simplest case
in detail. We begin by reviewing the effective action formulation of
the semi-classical eqs.~(\ref{eq:scE}) without regard to boundary
conditions or the state of the field. Thus, the equations in this
section are valid
for both the $\langle out | in \rangle$ and $\langle in | in \rangle$
formalisms.  We postpone to the next Section the introduction of the
CTP method which selects real and causal $\langle in | in \rangle$
expectation values. It is this latter form that must be used for the
linear response analysis.

The classical action for one scalar field (of arbitrary mass and
curvature coupling) is
\begin{equation} 
S_{\rm m}[\Phi, g]
= -\frac{1}{2}\int \d^4 x \;  \sqrt{-g}\, 
[ (\nabla_a \Phi)\, g^{ab}\, (\nabla_b \Phi) +m^2 \Phi^2+ \xi R\Phi^2] 
\; ,
\end{equation}
where $\nabla_a$ denotes the covariant derivative for the metric
$g_{ab}$, $\xi$ is the dimensionless curvature coupling, and $R$ is the
scalar curvature.  The path integral over the free scalar field $\Phi$
is Gaussian and may be computed formally by inspection, with the
result,
\begin{equation}
\int[{\cal D}\Phi]\exp \left({i\over \hbar}S_{\rm m}[\Phi, g]\right)
= \exp\left(-{1\over 2} {\rm Tr}
\ln G^{-1}[g]\right) \equiv 
\exp \left({i\over \hbar}S_{\rm eff}^{(1)}[g]\right)
\; ,
\label{eq:intM}
\end{equation}
where
\begin{equation}
G^{-1}[g] \equiv \sqrt{-g}\left(-\sq + m^2 + \xi R\right) 
\; ,
\end{equation}
is the inverse propagator of the scalar field in the background metric
$g_{ab}$, and the (generally non-local) functional
\begin{equation}
S_{\rm eff}^{(1)} [g]= {i\hbar \over 2}\,{\rm Tr} \ln G^{-1}[g] \equiv
{i\hbar \over 2} \int \d^4x'\,\sqrt{-g'} \ \langle x'\vert \ln G^{-1}[g]\vert
x'\rangle \label{eq:effM} 
\; ,
\end{equation}
may be regarded as the effective action due to the quantum effects of
the scalar field in this metric. It contains an explicit factor of
$\hbar$ and records the quantum effects of the free scalar field in
the arbitrary curved background spacetime $g_{ab}$.  No assumption
about the smallness of the metric deviations from flat spacetime or
any other preferred spacetime has been made.

The expectation value of the energy-momentum tensor of the quantum
matter field in this background can be formally obtained by the
variation,
\begin{eqnarray}
\langle T_{ab} (x)\rangle 
&=& -{2\over \sqrt{-g}} {\delta\over \delta g^{ab}} S_{\rm
eff}^{(1)}[g] \nonumber\\
&=& {-i\hbar\over \sqrt{-g}} {\rm Tr}\ G[g] 
{\delta G^{-1}[g] \over \delta
g^{ab}}  \equiv  {-i\hbar\over \sqrt{-g}} \int \d^4x'\sqrt{-g'} \int \d^4x''\
{\delta G^{-1}[g](x', x'') \over \delta g^{ab}(x)}\, G[g](x'', x')
\nonumber\\ & = & -i \hbar \sqrt{-g}
\;{\cal D}_{ab}\,G[g](x, x')\Big\vert_{x'=x} \; , 
\label{eq:Texp}
\end{eqnarray}
where ${\cal D}_{ab}$ is the second order operator (involving
covariant derivatives at $x$ and $x'$),
\begin{eqnarray}
{\cal D}_{ab}\,G[g](x, x')\Big\vert_{x'=x}&=&
\left[
\frac{1}{2}\nabla_a \nabla_b - \frac{1}{4} \eta_{ab} \sq 
- \frac{1}{2} m^2 \eta_{ab}
+ \xi \left(\eta_{ab} \sq - \nabla_a \nabla_b\right)
\right] G[g](x, x)
\nonumber
\\
&+& (- \delta_{a}^{c} \delta_{b}^{d}
+\frac{1}{2} \eta_{ab}\eta^{cd})
(\nabla'_c \nabla'_d)\,G[g](x, x')\Big\vert_{x'=x}
\; .
\label{eq:dab}
\end{eqnarray}
By Noether's theorem, this (unrenormalized) expectation value of
$T_{ab}$ is covariantly conserved, provided that the effective action
$S_{\rm eff}^{(1)}[g]$ is invariant under general coordinate
transformations.  However, $\langle T_{ab}\rangle$ is divergent
because of the singular nature of the limit $x' \rightarrow x$
in~(\ref{eq:Texp}), which requires a careful UV regularization and
subtraction procedure consistent with coordinate invariance, before a
finite renormalized value for its expectation value can be
defined~\cite{birrell-davies}.

In physical terms the UV regularization and renormalization procedure
mean that the theory is not strictly defined at arbitrarily short time
and distance scales. The lack of information about the physics at
those arbitrarily small scales may be absorbed into a finite number of
parameters in the effective low energy theory at larger scales. Since
the effective Lagrangian and energy-momentum tensor have canonical
scale dimension $n$ (in $n$ spacetime dimensions), the number of
parameters is given by the number of local coordinate invariant
scalars up to dimension $n$. In $n=4$ dimensions, these are the
parameters of the Einstein-Hilbert action plus the coefficients of the
two independent fourth order invariants $R^2$ and $C_{abcd}C^{abcd}$,
where $R$ is the scalar curvature and $C_{abcd}$ is the Weyl tensor,
respectively.  Thus, in order to renormalize the theory we require the
total low energy effective gravitational action,
\begin{equation} 
S_{\rm eff}[g] =
S_{\rm eff}^{(1)}[g] + {1\over 16\pi G_{_N}}\int \,  \d^4x\,
\sqrt{-g}\, (R-2\Lambda)  + {1\over 2} \int\, \d^4x\, \sqrt{-g} \, 
\left(\alpha C_{abcd}C^{abcd} + \beta R^2 \right) \; ,
\label{eq:effS} 
\end{equation}
with arbitrary dimensionless constants $\alpha$ and $\beta$.
Renormalization means that $G_{_N}$, $\Lambda$, $\alpha$, and $\beta$
are at first bare parameters, which may be chosen to depend on the UV
cutoff (introduced to regulate the divergences in the one-loop term
$S_{\rm eff}^{(1)}[g]$) in such a way as to cancel those divergences
and render the total action, $S_{\rm eff}[g]$, independent of the
cutoff. Hence the four parameters of the local geometric terms (up to
fourth order derivatives of the metric which are {\it a priori}
independent of $\hbar$) must be considered as parameters of the same
order as the corresponding divergent terms in $S_{\rm eff}^{(1)}[g]$,
which from~(\ref{eq:effM}) is first order in $\hbar$.  Formally, this
may be justified by considering $N$ identical copies of the matter
field, so that $S_{\rm eff}^{(1)}[g]$ is replaced by $NS_{\rm
eff}^{(1)}[g]$ and $G_{_N}^{-1}$, $\Lambda/G_{_N}$, $\alpha$, and
$\beta$ are rescaled by a factor of $N$. In this way all the terms in
eq.~(\ref{eq:effS}) are now of the same order in $N$ as $N \rightarrow
\infty$. 

This formal rescaling by $N$ is carried out at the level of the
generating functional of connected $p$-point vertices, $S_{\rm
eff}[g]$, (which are the {\it inverse} of $p$-point Green's functions)
rather than the Green's functions themselves. Therefore, it has the
net effect of resuming the quantum effects contained in the one-loop
diagrams of the matter field(s) to all orders in the metric $g_{ab}$.
The large $N$ expansion and its relationship to the standard loop
expansion have been extensively studied in both $\Phi^4$ theory and
electrodynamics (both scalar and spinor QED) in flat
space~\cite{CHKM}.  The QED case is most analogous to the present
discussion with the classical vector potential $A_{\mu}$ replaced by
the metric $g_{ab}$. The large $N$ approximation~(\ref{eq:effS}) is
also invariant under changes in the ultraviolet renormalization scale
(by definition of the UV cutoff dependence of the local counterterms
which cancel against those of the matter action), and is equivalent to
the UV renormalization group (RG) improved one-loop approximation.

It is the large $N$, RG improved one-loop approximation that is
necessary to derive the renormalized semi-classical
equations~(\ref{eq:scE}) with back-reaction, for only in such a
resummed loop expansion can the one-loop quantum effects of $\langle
T_{ab}\rangle$ influence the nominally classical background metric
$g_{ab}$.  As mentioned in the previous Section, in the ordinary
(unimproved) loop expansion the quantum fluctuations of the matter can
make at most small corrections to the background metric. The large $N$
approximation also preserves the covariance properties of the theory,
since it can be derived from an invariant action functional~(\ref{eq:effS}).
The divergences in $\langle T_{ab} \rangle$ are in one-to-one correspondence
with the local counterterms in the action $S_{\rm eff} [g]$, whose variations
with respect to $g_{ab}$ produce, in addition to the terms in the classical
Einstein equations, the fourth order tensors, 
\begin{mathletters} 
\begin{eqnarray} 
^{(1)}H_{ab}&&\equiv {1\over\sqrt{-g}}{\delta\over\delta g^{ab}} 
\int\, \d^4\,x\,\sqrt{-g}\, R^2 =
2 g_{ab}\sq R  -2\nabla_a\nabla_b R + 2 R R_{ab} - {g_{ab}\over 2}R^2 
\; ,  
\\
^{(C)}H_{ab}&&\equiv {1\over \sqrt{-g}}{\delta\over\delta g^{ab}}
\int\, \d^4\,x\,\sqrt{-g}\, C_{abcd}C^{abcd} = 4\nabla^c \nabla^d C_{acbd}
+ 2R^{cd}C_{acbd}
\; .
\end{eqnarray}
\label{eq:Hdef}
\end{mathletters}

\noindent
Hence the variation of the effective action~(\ref{eq:effS}) gives the
equations of motion for the spacetime metric (for zero expectation
value of the free scalar field $\Phi$),
\begin{equation} 
\alpha\ ^{(C)}H_{ab} +
\beta\ ^{(1)}H_{ab} + {1\over 8\pi G_{_N}}\left(G_{ab} + \Lambda
g_{ab}\right) =  \langle T_{ab}\rangle_{_R} \; ,  
\label{eq:scF}  
\end{equation}
where $\langle T_{ab}\rangle_{_R}$ is the renormalized expectation
value of the energy-momentum tensor of the scalar field, and all the
parameters are now understood to take finite renormalized values. In
order for the renormalized parameters to be defined unambiguously, we require
that any terms of precisely the form of the local geometric tensors on
the left hand side of~(\ref{eq:scF}), specified at an arbitrary but fixed
renormalization scale $\mu$, are removed from the expectation value on the right
side of (\ref{eq:scF}) by an explicit subtraction procedure at that scale
$\mu$. A concrete example of this subtraction procedure in flat spacetime is
given in Section IV.  

It is worth emphasizing that the UV renormalization of the
energy-momentum tensor and the covariant form of the equations of
motion~(\ref{eq:scF}) are justified by formal appeal to an underlying
covariant action principle~(\ref{eq:effS}), whose variation they are.
Although particular regularization and renormalization procedures,
such as non-covariant point splitting or adiabatic subtraction, may
break explicit covariance, the result must be of the
form~(\ref{eq:scF}), with a covariantly conserved $\langle
T_{ab}\rangle_{_R}$, or the procedure does not correspond to the
addition of local counterterms up to dimension $n=4$ in the effective
action, as required by the general principles of renormalization
theory. Thus, the renormalization of the effective
action~(\ref{eq:effS}) suffices in principle to renormalize the
equations of motion (\ref{eq:scF}) and {\it all} of its higher variations, a
fact we make use of in the next Section.

The large $N$ approximation is equivalent to a Gaussian path
integration for the quantum matter fields, in which the spacetime
metric and gravitational degrees of freedom have been treated as
$c$-numbers, coupled only to the expectation value of the
energy-momentum tensor through~(\ref{eq:scE}). Since the
energy-momentum tensor expectation value can be expressed as a
coincidence limit of local derivatives of the one-loop matter Green's
function $G[g](x,x)$ in the background metric $g_{ab}$ through eq.~(\ref{eq:Texp}), 
it requires solving the differential equation
$G^{-1}[g]\circ G[g] = 1$, or more explicitly
\begin{equation}
\left(-\sq + m^2 + \xi R\right) G[g](x, x') 
= {\delta^4(x,x')\over \sqrt{-g}} \; , 
\label{eq:grfn}
\end{equation}
concurrently with the semi-classical back-reaction
equation~(\ref{eq:scF}). It is the exact solution of this equation
without any perturbative re-expansion of $G[g]$, and the resulting
self-consistent solution of eq.~(\ref{eq:scF}) for the metric
$g_{ab}$, that constitutes the principal non-perturbative RG improved
feature of the large $N$ limit.

The equations of motion~(\ref{eq:scF}), which are the original
eqs.~(\ref{eq:scE}) modified by the additional terms required by the
UV renormalization of $\langle T_{ab}\rangle$, are fourth order in
derivatives of the metric. This feature, which is not present in QED
(but is a general feature of effective field theories which are
characterized by derivative expansions), has been the source of much
discussion in the
literature~\cite{horowitz,jordan-1,suen,flanagan-wald,simon,parker-simon}.  As
is well known from the general theory of differential equations, if the order
of the equations is changed by adding higher derivative terms, the solutions of
the modified equations fall into two classes, {\it viz.,} those that approach
the solutions of the lower order equations as the new parameters $\alpha, \beta
\rightarrow 0$, and those which become singular in that limit. The latter class
of solutions are not present in the lower order theory and correspond
physically to solutions which vary on Planck length and time scales (in order
for the higher derivative terms to be of the same order as the lower derivative
Einstein terms). There is clearly no experimental basis for taking these
solutions seriously (since they would predict that even empty flat space is
unstable to arbitrarily short length and time scale
perturbations)~\cite{horowitz,jordan-1,suen,flanagan-wald}.  Instead, the
modern framework of effective field theories suggests that we should regard the
Planck scale as the physical UV cutoff which defines the extreme limit of
possible validity of semi-classical gravity, and that we should confine our
attention to only those predictions of the theory which involve length scales
$\ell$ much greater than the Planck length $\ell_{\rm Pl}$ \cite{donoghue}. In
this regime, the effects of the higher order local terms in eq.~(\ref{eq:scF})
are suppressed by at least two powers of $\ell_{\rm Pl}/\ell$, provided the
solutions remain regular in the limit of vanishing renormalized coefficients
$\alpha$ and $\beta$ of the higher order terms. We are interested in this paper
primarily in defining a validity and stability criterion of the
semi-classical approximation at length scales $\ell \gg \ell_{\rm
Pl}$, and only comment briefly on the Planck scale solutions again in
the Discussion.

\section{CTP, Linear Response and the Stability Criterion}
\label{sec:3}

In this Section we present our criterion for the validity of the
semi-classical approximation which relies on a linear response
analysis. This analysis makes use of $\langle in|in\rangle$
expectation values which can be realized using the CTP
formalism~\cite{CTP}. We begin by reviewing a few details of this
formalism that are needed to derive the causal linear response
equation.

The desired $\langle in|in\rangle$ expectation values are obtained by
integrating the path integral~(\ref{eq:intM}) along a contour from the
initial time up to a late time in the future, and then backwards to
the initial time. This results in a doubling of the field variables
with a new CTP index (denoted by capital roman letters $A, B,
\dots$), running over the values $A=1,2$, that specify the forward or
backward part of the contour, respectively. After all manipulations
are performed the resulting expressions are evaluated by equating
field variables on the two contours.

In the CTP formalism the leading order effective action for the matter
fields is formally identical to eq.~(\ref{eq:effM}) with the
replacement,
\begin{equation}
G^{-1}[g] \rightarrow \left({\cal G}^{-1}[g]\right)_{AB} \equiv c_{AB} G^{-1}[g^A]
\; , 
\end{equation}
with ${\cal G}^{-1}[g]$ a $2 \times 2$ matrix in the CTP indices and
$c_{AB} = {\rm diag}(1,-1)$ the CTP metric. Thus, ${\cal
G}^{-1}_{11}[g]$ depends only on fields of type $1$, while ${\cal
G}^{-1}_{22}[g]$ depends only on fields of type $2$. The signs in the CTP
metric, $c_{AB}$, keep track of the direction of the time contour,
positively directed forward in time for field variables of the first
type, and negatively directed backward in time for field variables of
the second type.  The corresponding CTP effective action will be
denoted by ${\cal S}_{\rm eff}^{\rm CTP}$.  Performing the variation
of ${\cal S}_{\rm eff}^{\rm CTP}$ with respect to the first CTP
component of the metric variable, $g_{ab}^1$, is formally identical to
that of $S_{\rm eff}^{(1)}[g]$ in~(\ref{eq:Texp}), and gives $\langle
in|T_{ab}(x)|in\rangle$, the unrenormalized diagonal matrix element of the
energy-momentum tensor, which is real for hermitian $T_{ab}$.

The linear response equation can be obtained by expanding the CTP
effective action, in a functional Taylor series to one higher order,
around a given semi-classical geometry $g_{ab}$ that solves
eq.~(\ref{eq:scF}).  Writing
\begin{equation}
g_{ab} \rightarrow g_{ab} + h_{ab}
\; ,
\end{equation}
one finds that to second order in $h_{ab}$ the CTP effective action is
\begin{eqnarray}
{\cal S}_{\rm eff}^{\rm CTP} [g + h] &=& 
{\cal S}_{\rm eff}^{\rm CTP}[g] 
+ \int \d^4 x \; 
\frac{\delta{\cal S}_{\rm eff}^{\rm CTP}[g]}{\delta g_A^{ab}(x)}
\; h^A_{ab}(x)  
\nonumber \\ 
&+& 
\frac{1}{2} \int \d^4 x  \int \d^4 x'  
  \frac{\delta^2 {\cal S}_{\rm eff}^{\rm CTP}[g]}{\delta g_A^{ab}(x) 
\delta g_{cd}^B(x')}\;   h^A_{ab}(x) \; h^{B\,cd}(x') + \dots
\; ,
\label{eq:Sexpand}
\end{eqnarray}
where the first variation vanishes by~(\ref{eq:scF}).  Varying with
respect to $h^1_{ab}$ and then setting $h^1=h^2=h$ and $g^1 = g^2 =
g$, gives the linear response equation which is equivalent to the
first variation of the semi-classical Einstein equations, namely
\begin{eqnarray} 
&&
\delta \left[
\alpha\ ^{(C)}H_{ab} +
\beta\ ^{(1)}H_{ab} + {1\over 8\pi G_{_N}}\left(G_{ab} + \Lambda
g_{ab}\right) 
\right] = \delta \langle T_{ab}\rangle
\nonumber\\ 
&&=\frac{1}{4} {M_{ab}}^{cd} h_{cd}(x) -{1\over 2}\int \d^4
x'\sqrt{-g(x')}\, \Pi_{ab}^{({\rm ret})
\, cd}(x, x')\,h_{cd}(x')
\; ,
\label{eq:linres}
\end{eqnarray}
where
\begin{equation}
\Pi_{ab}^{({\rm ret})\, cd}(x, x') = \Pi_{ab}^{11\, cd}(x, x') +
\Pi_{ab}^{12\, cd}(x, x')
\; ,
\label{eq:piret} 
\end{equation}   
is the non-local connected, retarded polarization tensor and
${M_{ab}}^{cd}$ is the purely local part of the variation of $\langle
in|T_{ab}(x)|in\rangle$ at $x$.  We follow here the notation of
Ref.~\cite{mot-fd}, except for an opposite sign convention in the
definition of the energy-momentum tensor in eq.~(2.2) of that work.

To demonstrate that~(\ref{eq:piret}) is indeed the retarded
polarization tensor, we carry out the variation of the (unrenormalized) CTP
effective action for the scalar matter field explicitly, so that
\begin{eqnarray}
&&\qquad\qquad\qquad\delta \langle in\vert T_{ab}(x)\vert in\rangle = -{2\over
\sqrt{-g}}\int \d^4x' \frac{\delta^2 S_{\rm eff}^{\rm CTP}[g]}
{\delta g_1^{ab}(x)
\delta g_{cd}^B(x')} h^{B\,cd}(x') 
\nonumber
\\
&& \qquad \qquad \qquad\qquad =
-i\hbar{\rm Tr} \left({\cal G}[g]
\frac{\delta^2 {\cal G}^{-1}[g]}{\delta g_1^{ab}\delta g_{cd}^1}
h^{1\,cd}\right) + i\hbar {\rm Tr} \left({\cal G}[g]
 \frac{\delta
{\cal G}^{-1}[g]}{\delta g_1^{ab}} {\cal G}[g]
\frac {\delta {\cal G}^{-1}[g]}{\delta g_{cd}^B} 
h^{B\,cd}\right)
\nonumber\\ 
&& \qquad
= \frac{1}{4}{M_{ab}}^{cd} h_{cd}(x) + i\hbar  \int \d^4 x'
\sqrt{-g'}\left\{c_{11} {\cal D}_{ab} {\cal G}_{1A}[g](x,x') c_{AB}
{\cal D}^{cd} {\cal G}_{B1}[g](x',x)\right\} h_{cd} (x')\nonumber\\
&&\qquad = \frac{1}{4}{M_{ab}}^{cd} h_{cd}(x) + i\hbar  \int \d^4 x'
\sqrt{-g'}\left\{{\cal D}_{ab} {\cal G}_{11}[g](x,x'){\cal D}^{cd} {\cal
G}_{11}[g](x',x)  - {\cal D}_{ab} {\cal G}_{12}[g](x,x'){\cal D}^{cd} {\cal
G}_{21}[g](x',x) \right\} h_{cd} (x')
\; .  
\label{eq:secvar}
\end{eqnarray}
The definitions of the various components of the CTP matrix Green's
function of the scalar field are~\cite{CTP,CHKM,jordan-2}
\begin{eqnarray}
{\cal G}_{12}[g](x,x') &=& i\; \langle in|\Phi(x') \Phi(x) |in
\rangle  \equiv G_{<}(x,x')
\; ,
 \nonumber \\ 
{\cal G}_{21}[g](x,x') &=& i \; \langle in|\Phi(x) \Phi(x')|in \rangle  
\equiv G_{>}(x, x') = G_{<}(x',x)
\; ,
\nonumber \\
{\cal G}_{11}[g](x,x') &=& i \; \langle in|T [\Phi(x) \Phi(x')] |in \rangle
\equiv \theta (t,t') G_{>}(x,x') + \theta(t', t) G_{<}(x',x)
\; ,
\nonumber \\ 
{\cal G}_{22}[g](x,x') &=& {\cal G}_{11}^*[g](x,x')
= -\theta (t',t) G_{>}(x',x) - \theta(t, t') G_{<}(x,x')
\; .
\label{eq:grctp} 
\end{eqnarray} 
Hence suppressing momentarily the spacetime indices in the last line
of~(\ref{eq:secvar}), the CTP structure of the expression appearing there is
\begin{eqnarray} 
\theta (t,t') [G_{>}(x,x')]^2 + \theta(t', t)
[G_{<}(x',x)]^2 &-& G_{<}(x,x') \; G_{>}(x',x) = \theta (t, t')
\left\{[G_{>}(x,x')]^2 - [G_{<}(x',x)]^2\right\} \nonumber \\ &=&
-{1\over 2}\theta (t, t') \langle in \vert [\Phi^2(x),
\Phi^2(x')]\vert in \rangle \; ,
\end{eqnarray}
where we have used the definitions~(\ref{eq:grctp}) and the properties
of the Heaviside step function $\theta (t, t')$ for unequal arguments,
ignoring possible ambiguities at the coincident points $x=x'$.
Restoring the spacetime indices, we find that the non-local term of the
variation of $\langle T_{ab}\rangle_{_R}$ in~(\ref{eq:linres}) can be
written formally as
\begin{equation}
-\frac{1}{2}\int \d^4 x'\,\sqrt{-g'}\, \Pi_{ab}^{({\rm ret}) \, cd}(x,
x')\,h_{cd}(x') = -{i\hbar\over 2} \int \d^4 x'\, \sqrt{-g'}\,\theta
(t,t')  \; \langle in \vert [T_{ab}(x), T^{cd}(x')]\vert in
\rangle \; h_{cd}(x') 
\; ,
\label{eq:TTcor}
\end{equation}
which is real and causal.

This derivation is still formal because of the singular behavior of
the retarded polarization operator and possible ambiguities at the coincident
points, $x=x'$.  This singular behavior is related to the short distance
behavior of the formal expressions and their renormalization. The singular
behavior of commutators of physical currents and their various time ordered
products has been recognized for some time~\cite{photon}, and has been
discussed in the gravitational context in Ref.~\cite{mot-fd}. The
proper covariant definition of the singular functions requires
combining the retarded commutator with the first local (contact) term,
$\frac{1}{4} {M_{ab}}^{cd} h_{cd}(x)$ in~(\ref{eq:secvar}), in such a
way that the divergences in the sum of the two quantities can be
renormalized via the usual counterterms, namely exactly the same
counterterms at the level of the effective action which are necessary
to renormalize the semi-classical eqs. (\ref{eq:scF}) themselves.
Alternatively, one may calculate the time asymmetric part of the response
function, which is free of singularities in the limit $x \rightarrow x'$, and
{\it define} the renormalized time symmetric part of the full response function
(including the local contact terms) by a covariant regularization and
renormalization procedure, which gives unique answers up to finite
redefinitions of the coefficients $\alpha$ and $\beta$ in the fourth
order renormalized effective action. It is this latter procedure which
we carry out explicitly by means of a dispersion integral, after
Fourier transforming~(\ref{eq:TTcor}) in the flat space example
provided in the next Section.

The linearized fluctuation $h_{ab}(x)$ obeys an integro-differential
equation~(\ref{eq:linres}) in which the integral depends only on the
past of $x$, due to the causal boundary conditions, and which involves
the two-point correlation function of the matter energy-momentum
tensor.  According to the general principles of linear response
analysis, this retarded correlation function is evaluated in the
background geometry of the leading order solution of the
semi-classical equations~(\ref{eq:scF}).

The polarization operator, $\Pi_{ab}^{({\rm ret})\, cd}(x, x')$, is
determined by the second variation of the same effective action that
determines the energy-momentum tensor, and it also obeys the same
covariant conservation law,
\begin{equation} 
\nabla^a \Pi_{ab}^{({\rm ret})\, cd}(x, x') 
= \nabla'_{c} \Pi_{ab}^{({\rm ret})\, cd}(x, x') = 0\,,\qquad {\rm for}
\qquad x \neq x'
\; .
\end{equation}
The equations~(\ref{eq:linres}) are covariant in form and therefore
are non-unique up to linearized coordinate (gauge) transformations
\begin{equation}
\delta g_{ab} \rightarrow \delta g_{ab} + \nabla_a X_b + \nabla_b X_a
\; ,
\label{eq:gauge}
\end{equation}
for any vector field $X_a$. Singular gauge transformations in the
initial data for $\delta g_{ab}$ are excluded, and some
care is required to decide whether time dependent linearized gauge
transformations which grow in time without bound are allowed or not.
Since the action principle is fundamental to the present approach, any
transformation of the form~(\ref{eq:gauge}), for which the
action~(\ref{eq:effS}) is not invariant (due to boundary or surface
terms), is not a true invariance and should be excluded from the set
of allowable gauge transformations of the linear response
equations~(\ref{eq:linres}).

We now state our stability criterion for the semi-classical
approximation. A necessary condition for the validity of the large $N$
semi-classical equations of motion~(\ref{eq:scF}) is that the linear
response equations~(\ref{eq:linres}) should have no solutions with
finite non-singular initial data for which any linearized gauge
invariant scalar quantity grows without bound.  Such a quantity must
be constructed only from the linearized metric perturbation $h_{ab}$
and its derivatives, and it must be invariant under allowed gauge
transformations of the kind described by eq.~(\ref{eq:gauge}).
 
The existence of any solutions to the linear response equations with
unbounded growth in time, that cannot be removed by an allowed
linearized gauge transformation~(\ref{eq:gauge}), implies that the
influence of the growing gravitational fluctuations on the
semi-classical background geometry are large, and must be taken into
account in the evolution of the background itself. That is to say, if
the gravitational fluctuations around the background grow, even if
they were initially small, then the leading order semi-classical
equations~(\ref{eq:scF}), which neglect these fluctuations, must
eventually break down.

\section{Stability of Flat Spacetime}
\label{sec:4}

Flat spacetime is a solution of the semi-classical Einstein equations
for vanishing expectation value of $T_{ab}$ and cosmological term,
with the quantum matter field in its Lorentz invariant vacuum ground
state. This is the simplest solution of the semi-classical
eqs.~(\ref{eq:scF}) to which we can apply our validity criterion, and
for which the polarization operator, $\Pi_{ab}^{({\rm ret})\, cd}$,
can be evaluated in closed form. In addition to illustrating the
application of the criterion to a well-defined specific case, the
analysis of the normal modes which solve the linear response
eq.~(\ref{eq:linres}), will permit us to reach a definitive conclusion
on the stability of Minkowski spacetime to quantum perturbations on
distance and time scales far larger than the Planck scale.

The linear response equation~(\ref{eq:linres}) around a Minkowski
background, $\eta_{ab}$, can be decomposed into scalar, vector and
tensor components according to the decomposition and projection operators
defined in Appendix A.  The variations of the local tensors appearing
on the left hand side of eq.~(\ref{eq:linres}) are
given by eqs.~(\ref{eq:wre}). Thus, eq.~(\ref{eq:linres}) around flat
space may be written in the form,
\begin{eqnarray}
&&\left[\alpha \sq^2 - {1\over 16\pi G_{_N}} \sq\right] P_{ab}^{(T)\,
cd}h_{cd}(x) + \left[6\beta \sq^2 + {1\over 8\pi G_{_N}} \sq\right]
P_{ab}^{(S)\, cd}h_{cd}(x)\nonumber\\
&& \qquad = \frac{1}{4} M_{ab}^{\ \ cd}
h_{cd}(x)-{1\over 2}\int \d^4 x'\,\Pi_{ab}^{({\rm ret})\, cd}(x,x')\,h_{cd}(x')
\; .
\label{eq:lrflat}
\end{eqnarray}
The non-local vacuum polarization tensor [right hand side of
eq.~(\ref{eq:lrflat})] can be decomposed into exactly the same two
scalar and tensor projections, (see both Appendix A and B),
$P_{ab}^{(T)\, cd}\Pi^{(T)\,({\rm ret})} + P_{ab}^{(S)\,
cd}\Pi^{(S)\,({\rm ret})}$, and a K\"all\'en-Lehmann spectral
representation~\cite{leb,FW,kal-leh} given for the Fourier transform
of each of these two gauge invariant scalar functions,
\begin{equation} 
\Pi^{(i)\, ({\rm ret})}(k^0, \vec k) = \int_{-\infty}^{\infty} 
{\d\omega\over2\pi} 
\ {\sigma^{(i)}(\omega, \vec k) \over \omega - k^0 - i\epsilon}  \; ,
\qquad i = T,S 
\; .
\label{eq:ac} 
\end{equation} 
This is a form of Cauchy's theorem for 
\begin{equation}
{\rm Im}\left[ \Pi^{(i)\, ({\rm ret})}(k^0, \vec k)\right]
 = {1\over 2} \sigma^{(i)}(k^0,
\vec k) \equiv \pi \rho^{(i)} (s)\; {\rm sgn}(k^0)
\; , 
\qquad s = (k^0)^2 - |\vec k|^2
\; .
\label{eq:cauchy}
\end{equation}
In Fourier space the non-local real convolution in
eq.~(\ref{eq:lrflat}) becomes a simple multiplication with
$h_{cd}(k)$. Its real part is even and its imaginary part odd under
time reversal, which is taken into account by the sgn$(k^0)$ function
in~(\ref{eq:cauchy}), and which is a consequence of the causal, retarded
boundary conditions of the CTP $\langle in|in\rangle$ formalism.

Since the purely local term, $M_{ab}^{\ \ cd}$, is time reversal
invariant, it does not contribute to the imaginary part of the
dispersion relation~(\ref{eq:ac}) of $\Pi^{(i)(ret)}(k^0, \vec k)$ which is
finite and well-defined. The proper definition of the local term is connected
with the renormalization procedure needed to reconstruct the real part
of the Fourier transform of~(\ref{eq:lrflat}) from its imaginary part,
and in fact, the dispersion integral in eq.~(\ref{eq:ac}) does {\it not}
exist due to the large $k$ behavior, {\it viz.} $k^4$, of the Lorentz
invariant spectral functions $\rho^{(i)}(s=-k^2)$.  This divergent
behavior of the unrenormalized dispersion integral~(\ref{eq:ac}) as
$k^2 \rightarrow \infty$, is nothing but the ambiguities of the
coincident limit $x \rightarrow x'$ in the retarded polarization
function in a different guise. The divergent terms are proportional to
$\delta^4(x,x')$ and up to four derivatives thereof, which by Lorentz
invariance must be of exactly the same form as the local terms on the
left hand side of the linear response eq.~(\ref{eq:linres}). Thus,
these divergences, as well as the local term $M_{ab}^{\ \ cd}$, can be
handled by the same renormalization procedure needed to define the
expectation value of the dimension four operator $T_{ab}$
in~(\ref{eq:scE}), namely by subtraction of the allowed covariant
counterterms up to dimension four. In the flat space dispersion
integral~(\ref{eq:ac}), this is easily accomplished by subtracting the
first three terms in its Taylor series expansion around $k^2 =0$, and
defining the renormalized real part of the retarded correlation
function by
\begin{eqnarray}
{\rm Re}\left[\Pi^{(i)\, ({\rm ret})} (k^2)\Big\vert_{\rm R} \right] &\equiv &
{\rm Re} \left\{ \Pi^{(i)\, ({\rm ret})}(k^2) - \Pi^{(i)\, ({\rm ret})}(0)
- k^2 \left[{\partial \Pi^{(i)\, ({\rm ret})}\over \partial
k^2}\Big\vert_{k^2=0}\right] - {(k^2)^2\over 2} \left[{\partial^2 \Pi^{(i)\,
({\rm ret})}\over \partial (k^2)^2}\Big\vert_{k^2=0}\right]\right\}\nonumber\\
&& \qquad = -(k^2)^3\ {\cal P} \int_0^{\infty} {\d s\over s^3} 
{\rho^{(i)}(s)\over s + k^2} 
\; ,
\label{eq:renpi} 
\end{eqnarray}
where $\cal P$ denotes the principal part prescription for the
integral when $k$ is timelike ($-k^2 = s$). The subtractions do not
affect the time odd imaginary part of the retarded polarization
function. The integral over $s$ in the real part is now well-defined
and UV finite, and may even be computed in terms of elementary
functions, in the case of a scalar field of arbitrary mass $m >0$ and
curvature coupling $\xi$. The details of this calculation are given in
Appendix B. The three subtractions in~(\ref{eq:renpi}) correspond physically
to renormalizing the coefficients of the cosmological constant ($\Lambda$),
Newton's constant ($1/G_{_N}$), and the coefficients of the fourth order terms
($\alpha$ and $\beta$). These are of order $(k^2)^0$, $k^2$  and $(k^2)^2$
respectively. The renormalized values of these parameters at $k^2 =0$ are what
appear then on the left hand side of (\ref{eq:scF}). The singular local term, 
$M_{ab}^{\ \ cd}$ in (\ref{eq:lrflat}) is effectively removed by these
subtractions in flat space as well, so that the entire linear response eq.
becomes well-defined and covariant. Indeed had we performed the renormalization
at the level of the effective action directly then it would be clear that
no local term ambiguities appear in the renormalized equations.

Since the projections onto scalar and tensor modes are linearly
independent (in fact, orthogonal), the coefficients of the two
projection operators must satisfy the linear response relation
separately. Transferring the polarization part to the left hand side
of~(\ref{eq:lrflat}) and taking account of the renormalization just
described, yields two independent dispersion formulae, namely,
\begin{mathletters}
\begin{eqnarray}
&& k^2 \left[2 \alpha k^2 + {1\over 8\pi G_{_N}} + (k^2)^2 \int_0^{\infty}
{\d s\over s^3}  {\rho^{(T)}(s)\over (s + k^2 - i\epsilon\ {\rm sgn}
(k^0))}\right] =0
\; ,
\label{eq:lrten}\\ 
&& k^2 \left[ 12\beta k^2 - {1\over 4\pi G_{_N}} + (k^2)^2
\int_0^{\infty} {\d s\over s^3}  {\rho^{(S)}(s)\over (s + k^2 - i\epsilon\ {\rm
sgn} (k^0))}\right] =0 
\; .
\label{eq:lrsca}
\end{eqnarray}
\end{mathletters}
The two spectral functions are calculated explicitly for the free
scalar field with arbitrary mass and curvature coupling in Appendix
B. For this case, the spectral functions have support only when $s >
4m^2$, which corresponds to the two particle threshold for timelike
gravitational fluctuations. However, some conclusions can be drawn
from the two dispersion relations above using only the fact that both spectral
functions are positive for both transverse tensor and scalar gravitational
perturbations of flat space. 

Let us examine first the tensor dispersion relation. It clearly is
always satisfied by $k^2 =0$. This solution corresponds to the
physical, transverse linearized gravitational waves propagating in a
flat space background. The coefficient of $k^2$ at $k^2=0$ is
unchanged from the classical value by the quantum parameter $\alpha$
and vacuum polarization corrections.  Therefore, these linearized
gravitational waves carry the same energy density in the
semi-classical approximation as they do in the classical Einstein
theory.

Next we may examine the interior of the brackets to determine if there
are any other solutions to the tensor linear response
equations. Solutions with $k^2 = -(k^0)^2 + |\vec k|^2 >0$ correspond
to unstable modes with imaginary frequencies, since we can always
consider these modes in a frame where $\vec k = 0$. When $k^2 >0$ the
$-i\epsilon$ prescription is not needed, and $\epsilon$ may be set to
zero. Thus, by making use only of the positivity of $\rho^{(T)}$, 
and assuming that $\alpha$ is non-negative we
observe that the bracket is strictly positive for $k^2 >0$, provided
\begin{equation}
16\pi \alpha G_{_N} k^2 + 1 > 0
\; .
\label{eq:lesspl}
\end{equation}
If $\alpha \ge 0$ this is always satisfied, and indeed 
this constraint on $\alpha$ is required by positivity of the energy density
$-2^{(C)}H_{00}$ corresponding to the fourth order $C_{abcd}C^{abcd}$ term in
the action. This demonstrates that there are no unstable transverse tensor
perturbations of flat spacetime for $\alpha \ge 0$. Since this conclusion
relies only on the positivity of the spectral function $\rho^{(T)}(s)$, it
requires only causality, a bounded Hamiltonian, and a well-defined positive
Hilbert space norm for the quantum matter theory. Hence it is valid much more
generally than for the specific scalar field example.

Going further, we may inquire as to the existence of additional stable
solutions characterized by propagating tensor wave modes with timelike
$k$ ($k^2 <0$).  The bracket in~(\ref{eq:lrten}) vanishes if 
\begin{equation} 
16\pi \alpha G_{_N}k^2 + 1 + 8\pi G_{_N} k^2 F^{(T)} = 0
\; ,
\label{eq:tendis}
\end{equation}
where 
\begin{equation}
F^{(T)} \equiv k^2 \int_0^{\infty} {\d s\over s^3} 
{\rho^{(T)}(s)\over (s + k^2 - i\epsilon\ {\rm sgn} (k^0))}
\; ,
\end{equation}
is a dimensionless function of $k^2/m^2$ (and the sign of $k^0$),
given explicitly for the case of a scalar field by~(\ref{eq:tenF}). It
is clear that if this function remains bounded for all $k^2/m^2$, the
equality~(\ref{eq:tendis}) can never be satisfied for $16\pi G_{_N}
|k^2| \ll 1$, since both the third (polarization) term and first
($\alpha$) term can never be of order unity. In fact, from the
explicit form of $F^{(T)}$ for a scalar field, given by eq. (\ref{eq:tenF}),
\begin{equation}
{\rm Re}\left[ F^{(T)}\left({k^2\over m^2}\right)\right]
 \rightarrow {1\over 960\pi^2}
\ln
\left({\vert k^2\vert\over m^2} \right)\qquad {\rm as} \qquad \vert k^2\vert
\rightarrow \infty
\; ,
\label{eq:FT} 
\end{equation}
so that the function does grow without bound, but only
logarithmically. Hence the relation~(\ref{eq:tendis}) cannot be
satisfied except at $k^2$ approaching $G_{_N}^{-1}$, provided $m^2
>0$.  If $\alpha <0$ then the preceding analyses for $k^2 >0$ and
$k^2<0$ interchange roles, with the conclusion unchanged. Thus, there are
no tensor mode solutions of the linear response eq.~(\ref{eq:linres}) on length
scales much larger than the Planck length for a massive field theory around
flat space, other than the usual linearized gravitational waves of the
classical theory. On physical grounds one must expect this result to hold for
any quantum matter field polarization tensor of finite mass obeying the same
general properties of our scalar field example.

The logarithmic divergence in the response function $F^{(T)}$ 
when $k^2 \rightarrow \infty$ is a consequence
of the large $s$ (UV) behavior of the spectral function proportional to $s^2$,
and is generic, with only the value of the finite coefficient of the logarithm
dependent on the matter content. However the appearance of $m^2$
in the lower limit of the logarithm is a result of our definition of
renormalized $\alpha$ parameter at $k^2 =0$, which allows no other
scale to appear in the logarithm. This definition is no longer tenable in the
zero mass limit. Instead, one may renormalize the parameter at an arbitrary
non-zero value of $k^2 = \mu^2$, related to our previous definition by
\begin{equation}  
\alpha (\mu^2) = \alpha + {1\over 1920\pi^2} 
\ln \left( {\mu^2\over m^2}\right)
\; . 
\label{eq:alpmu} 
\end{equation} 
Then, $\mu^2$ replaces $m^2$ in the logarithm of eq.~(\ref{eq:FT}),
and the $m^2 \rightarrow 0$ limit may be taken safely by maintaining
$\alpha(\mu^2)$ finite. The conclusions about the absence of solutions
to the transverse linear response equation (\ref{eq:lrten}) at length scales
much greater than the Planck length remains unchanged.

Turning now to the scalar component of the linear response equation,
(\ref{eq:lrsca}), we note the opposite sign in the coefficient of the
$1/G_{_N}$ term, which is the well-known negative metric sign of the conformal
factor in the Einstein-Hilbert action. From several different
analyses~\cite{har-sch,mazmot} it is known that there are no physical wavelike
scalar excitations of flat space in either the classical or
semi-classical theory. Physically the reason is that the conformal
factor is constrained by the diffeomorphism invariance.  This implies
that the expansion of the gravitational action to second order in the
metric perturbations about flat space, $h^{(S)\, ab} k^2 h^{(S)}_{ab}$ should be
treated as proportional to $|\tilde \chi|^2$, where $\tilde\chi$ is the Fourier
transform of a new scalar field variable with no kinetic term in the
Einstein-Hilbert action. This redefinition may be understood as required also
in the covariant path integral treatment of the linearized gravitational
fluctuations around flat space~\cite{mazmot}. The net effect of either the
covariant or canonical analysis is to remove the overall $k^2=0$
propagating solution from the scalar sector, as it is not dynamically allowed by
the constraints. The possibility of a finite number of non-propagating
(global) modes in the scalar sector at $k=0$ cannot be excluded by this
argument, and must be investigated by imposing appropriate boundary conditions
at spatial infinity. We do not treat this possibility in this paper.

Finally, the analysis of the expression within the brackets
of~(\ref{eq:lrsca}) shows that there are no scalar mode solutions
(stable or otherwise) with $4\pi G_{_N} |k^2| \ll 1$, for exactly the
same reason as in the tensor case, notwithstanding the sign change in
the Einstein term. In the scalar case the explicit form of the response
function (\ref{eq:scaF}) yields
\begin{equation}
{\rm Re}\left[F^{(S)}\left({k^2\over m^2}\right)\right]
 \rightarrow {1\over 96\pi^2}
(1-6\xi)^2 \log\left({\vert k^2\vert\over m^2} \right)\qquad {\rm as} \qquad
\vert k^2\vert \rightarrow \infty 
\; .
\label{eq:FS}
\end{equation}
which shows that the linear response eq. cannot be satisfied for either
sign of $k^2$ unless $G_{_N} |k^2|$ becomes of order unity. This
result was first obtained in Ref.~\cite{mazmot} for the case of a massive, but
conformally coupled field with $\xi = 1/6$. In the conformal case, the large
$s$ behavior of the scalar spectral function $\rho^{(S)}$ is much less severe,
as is clear from its explicit form given in eq.~(\ref{eq:rho-S-con}), and
only two subtractions suffice. This corresponds to only a finite
renormalization of the $\beta R^2$ term in the effective action. Ignoring the
$\beta$ term completely, the only twice subtracted
dispersion formula gives then \begin{equation}
1 + 4\pi G_{_N} k^2 \int_0^{\infty} {\d s\over s^2}  {\rho^{(S)}(s)\over (s +
k^2 - i\epsilon\, {\rm sgn} (k^0))} = 0
\; ,
\label{eq:lds}
\end{equation}
in the scalar sector. This form of the linear response equation was
used in Ref.~\cite{mazmot} to demonstrate the stability of flat space
to scalar (conformal) fluctuations in the infrared limit, {\it i.e.,}
on wavelengths far larger than the Planck length. Indeed,
substituting the explicit form of $\rho^{(S)}$, eq.~(\ref{eq:rho-S})
with $\xi =1/6$, into~(\ref{eq:lds}), shows that it is identical with
eq.~(4.13) of~\cite{mazmot} with $2m^4 \rho (s)/3$ of that reference
equal to $\rho^{(S)}(s)$ here.

When $\xi \neq 1/6$ this argument cannot be used since the scalar
spectral function behaves as $s^2$ for large $s$ and the twice
subtracted dispersion integral in (\ref{eq:lds}) diverges. However,
analysis of the fully subtracted response function behavior in (\ref{eq:FS})
above leads to the same result. As in the tensor case the logarithmic
growth with $|k^2|$ is generic, with only the coefficient of the
logarithm depending on the matter content, possibly vanishing in
some special cases such as $\xi = 1/6$. Also as in the tensor case the
lower limit of the logarithm can be made finite in the $m^2 \rightarrow 0$
limit by redefining the renormalized $\beta$ coefficient at a finite $\mu^2
\neq 0$ analogously to eq.~(\ref{eq:alpmu}). One obtains then the same
conclusion as for the tensor case, namely that there are no new
solutions  of the linear response equations, stable or unstable, far from the
Planck regime, despite the opposite sign of the classical Einstein term in the
scalar sector.

\section{Discussion}
\label{sec:5}  

We have presented a criterion for the validity of the semi-classical
approximation for gravity that involves solving the linear response
equation~(\ref{eq:linres}), to determine the stability of solutions to
the semi-classical equations~(\ref{eq:scF}). If, for a given state and
background geometry that solves the semi-classical equations, one or
more solutions to the linear response equations experience unbounded
growth in a gauge invariant sense, then the semi-classical
approximation is not valid for that particular geometry, at least not
for that particular state. Clearly this is a necessary, though perhaps not
a sufficient condition for the validity of the approximation.

As discussed in the introduction, various methods have been suggested
previously to test the validity of the semiclassical approximation by
making use of the two-point correlation function for the
energy-momentum tensor. The linear response criterion provides a
natural and well-defined way for this two-point correlation function
to enter into the determination of the validity of the semi-classical
approximation.  Further, linear response involves quantities that lie
entirely within the semi-classical approximation itself, since the
polarization tensor is computed on the semi-classical background
geometry.  The large $N$ method, augmented by the causal CTP
formulation of the effective action, provides a well-defined framework
for applying the validity criterion, which is equivalent to a
stability criterion for the semi-classical solution.

In the covariant effective action formulation, it is clear that the UV
renormalization counterterms are the same as those needed to define
the semi-classical approximation itself and that there are no
state-dependent divergences. Although the matter energy-momentum
tensor correlator by itself suffers from possible ambiguities at
coincident points, these are removed by a proper covariant
regularization and renormalization procedure, which ties these
divergences in the correlator at $x=x'$ to counterterms in the purely
local gravitational effective action.  The resulting combination of
{\it all} terms in eq.~(\ref{eq:linres}) then becomes well-defined.
 
We have illustrated the use of the stability criterion with the simple
example of a quantized scalar field with arbitrary mass and curvature
coupling in the vacuum state of Minkowski spacetime. In this case it
is possible to carry out the analysis to completion and show that flat
space is stable in the infrared limit. There are no propagating solutions to
the linear response eqs.~(\ref{eq:linres}) around flat space, except the
usual transverse, traceless gravitational wave excitations of the
classical Einstein theory, provided we restrict ourselves to solutions
with $4\pi G_{_N} |k^2| \ll O(1)$. The exact finite number of order unity on
the right hand side of this inequality determines the values of $k^2$ for which
new Planckian solutions and instabilities will appear.  Its value depends on
both the matter theory and the values of the renormalized coefficients $\alpha$
and $\beta$ of the fourth order terms. The existence of such growing modes,
which violate the validity criterion proposed here,  informs us not that flat
spacetime is unstable, but only that the quantum fluctuations of the geometry
should be included in some consistent way at short scales. It is then the
semi-classical notion of flat space as a  pseudo-Riemannian manifold endowed
with a smooth metric down to  arbitrarily short length scales that is breaking
down at the Planck scale. The semi-classical approximation which does not
incorporate the effects of these quantum fluctuations on the mean geometry is
certainly not valid in the Planck regime. However, the semi-classical
approximation contains the signal of its own breakdown at such short scales
according to the validity criterion, and leads to a completely satisfactory
stability of flat space for all local perturbations obeying the
inequality~(\ref{eq:lesspl}).

That empty flat space with quantum matter in its vacuum ground state
should be stable, and quantum gravitational effects negligible
excluding at the Planck scale, is hardly surprising. It does mean that
the predictions of the semi-classical approximation at least are not
in complete disagreement with observations in this case. In addition
to providing an explicit example of how to handle the energy-momentum
tensor correlation function by standard renormalization methods to
obtain well-defined answers, working out this case in detail also
provides an important clue as to how the validity criterion may fail
to be satisfied in more interesting cases.  What is required is simply
that the polarization tensor of the matter fluctuations become
singular, {\it i.e.,} large and unbounded, in some region.  Only in
this way can the natural suppression of $G_{_N}k^2 \ll 1$ in flat
space be overcome. A non-trivial example, where new modes may be
expected, is a finite temperature quantum matter field in an
Einstein-de Sitter model~\cite{rebhan}. It would be interesting to
apply the validity criterion proposed here to a consistent solution of
the semi-classical Einstein equations possessing thermal matter.

Further important and interesting examples to which the criterion may
be applied are solutions with event horizons, such as Schwarzschild
and de Sitter spacetimes, as well as more general cosmological
solutions of the semi-classical equations. If the linearized solutions
of~(\ref{eq:linres}) show any growing modes, due to a singular
behavior of the polarization tensor and/or the existence of
non-trivial global modes in such cases, then one would be led to the
conclusion that including gravitational fluctuations beyond the leading
order semi-classical approximation in some form would be required. A number of
different arguments lead to the conclusion that de Sitter spacetime is not the
stable ground state of a quantum theory of gravity with a cosmological
term~\cite{mot-des}. In fact, the two-point correlation function of the
energy-momentum tensor for a scalar field was estimated in~\cite{mot-fd}, and
argued to contribute to a gauge invariant growing mode on the horizon time
scale.  This proposition could be tested by a detailed calculation of the
two-point correlation function of the energy-momentum tensor and the solutions
of the linear response equations~(\ref{eq:linres}) in de Sitter space.

A second important application of the criterion is to black hole
spacetimes. Ever since the discovery of black hole radiance, it has
been recognized that the quantum behavior of black holes is
qualitatively different from the classical analogs at {\it long}
times, since semi-classical black holes decay at late times, while
classical black holes are stable. In the Hartle-Hawking
state~\cite{harhaw} one can construct a static solution to the
semi-classical equations~(\ref{eq:scF}) that is quite close to the
classical one near the horizon~\cite{york,hky,ahwy}. On thermodynamic
grounds this state is expected to be
unstable~\cite{unstable}. However, the stability of this
self-consistent solution has not been investigated in a dynamical
approach. The validity criterion proposed in this paper provides a
clear dynamical principle for the stability or instability of the
self-consistent solutions in both the black hole and de Sitter cases.

\acknowledgements P.\ R.\ A.\ and C.\ M.-P.\ would like to thank T-8,
Los Alamos National Laboratory for its hospitality.
P.\ R.\ A.\ would like to thank J. Donoghue, L. Ford, B.L. Hu, and N. Phillips
for helpful conversations.  E. M. thanks P. O. Mazur for reading the manuscript
prior to publication. This work was supported in part by grant numbers
PHY-9800971 and PHY-0070981 from the National Science Foundation. It was also
supported in part by contract number W-7405-ENG-36 from the Department of
Energy.

\appendix
\section{Tensor Decomposition and Spectral Representation in Flat
Space}

An arbitrary metric perturbation, $h_{ab}$, around $n$ dimensional
flat space, $\eta_{ab}$, can be decomposed in the following way:
\begin{equation}
h_{ab} = h_{ab}^{\perp} + \partial_a v_b^{\perp} + \partial_b v_a^{\perp}
+ \left(\partial_a\partial_b -{1\over n}\eta_{ab} \sq\right) w + 
{\eta_{ab}\over n} h
\; ,
\end{equation}
where $h_{ab}^{\perp}$ is transverse and traceless with respect to the
flat metric $\eta_{ab}$
\begin{equation}
\partial^a h_{ab}^{\perp} = 0 = \partial_b h_{ab}^{\perp}
\; ,
\; \; \;
{\rm and} \; \; \;
\eta^{ab}h_{ab}^{\perp}=0
\; ,
\end{equation}
and $v_a^{\perp}$ is transverse
\begin{equation}
\partial^a v_a^{\perp} = 0
\; .
\end{equation}
By taking partial derivatives and traces and using these defining
properties, the various terms in the decomposition can be isolated
successively, {\it viz.}
\begin{eqnarray}
h &=& \eta^{ab}h_{ab}
\; ,
\\
w &=& -{1\over n-1} \sq^{-1} \left(\eta^{ab}-
n\sq^{-1}\partial^a \partial^b\right) h_{ab}
\; ,
\\
v_a^{\perp} &=& \sq^{-1}\left(\delta_a^c - \sq^{-1} \partial_a\partial^c\right)
\partial^d h_{cd}
\; ,
\\
h_{ab}^{\perp} &=& \left[
\delta_a^c \delta_b^d - {1\over n-1}\eta_{ab}\eta^{cd}
+{2\over n-1}
\sq^{-2}\partial_a\partial_b\partial^c\partial^d
\right.\nonumber\\
&+& \left.
\sq^{-1} \left( - \delta_a^c \partial_b\partial^d - \delta_b^c
\partial_a\partial^d + {1\over n-1} \eta_{ab} \partial^c\partial^d
+ {1\over n-1} \eta^{cd} \partial_a\partial_b\right)
\right]h_{cd}
\; ,
\end{eqnarray}
where $\sq^{-1}$ denotes the propagator inverse of $\sq \equiv \eta^{ab}
\partial_a\partial_b$.

Under an infinitesimal coordinate (gauge) transformation,
\begin{equation}
h_{ab} \rightarrow h_{ab} + \partial_a X_b + \partial_b X_a
\; ,
\end{equation}
the change in $h_{ab}$ can be absorbed into a redefinition of the
various components of the decomposition according to
\begin{eqnarray}
&&h \rightarrow h + 2 \sq Y
\; ,
\nonumber\\
&&w \rightarrow w + 2 Y
\; ,
\nonumber\\
&&v_a^{\perp} \rightarrow v_a^{\perp} + X_a^{\perp}
\; ,
\nonumber\\
&&h_{ab}^{\perp} \rightarrow h_{ab}^{\perp}
\; ,
\end{eqnarray}
where $X_a$ has been decomposed into its transverse and longitudinal
parts, as 
\begin{eqnarray}
X_a = X_a^{\perp} + \partial_a Y \; ,
\; \; \; {\rm with} \; \; \; 
\partial^a X_a^{\perp} = 0
\; .
\end{eqnarray}
{From} these transformations we observe that the transverse, traceless
tensor $h_{ab}^{\perp}$ and the linear combination of scalars
\begin{equation}
h- \sq w = {n\over n-1} \left( \eta^{cd} - \sq^{-1}\partial^c
\partial^d\right) h_{cd}
\; ,
\end{equation}
are invariant under infinitesimal coordinate transformations. Hence we
may define the projections onto the scalar (spin-$0$) and transverse,
traceless tensor (spin-$2$), gauge invariant terms in the general
decomposition of the symmetric tensor perturbation $h_{ab}$ by
\begin{mathletters}
\begin{eqnarray}
h_{ab}^{(S)} &\equiv& {1\over n-1} \left( \eta_{ab} -
\sq^{-1}\partial_a  \partial_b\right) \left( \eta^{cd} -
\sq^{-1}\partial^c \partial^d\right)h_{cd}\nonumber\\ 
&=& {1\over n} \left( \eta_{ab} - \sq^{-1} \partial_a
\partial_b\right)(h-\sq w) \equiv P^{(S) \, cd}_{ab} h_{cd}
\; ,
\\ h_{ab}^{(T)} &\equiv& h_{ab}^{\perp} \equiv P^{(T) \, cd}_{ab} h_{cd}
\; .
\end{eqnarray} 
\end{mathletters}

\noindent The remaining terms in the decomposition contain all the gauge
dependence. We denote the vector perturbation [containing both
transverse (spin-$1$) and longitudinal (spin-$0$) components] by
\begin{eqnarray}
h_{ab}^{(V)} &=& \sq^{-1} \left(\delta_a^c \partial_b + \delta_b^c \partial_a
- \sq^{-1} \partial_a\partial_b\partial^c\right)
\partial^dh_{cd}
\equiv P^{(V) \, cd}_{ab} h_{cd}
\; .
\end{eqnarray}
Thus, the general symmetric tensor metric perturbation can be written
as the sum of three projected components
\begin{equation}
h_{ab} = h_{ab}^{(S)} + h_{ab}^{(V)} + h_{ab}^{(T)}
= \sum_{i = S, V, T} P^{(i)\, cd}_{ab} h_{cd}
\; .
\end{equation}
The three projectors are orthonormal, {\it i.e.,}
\begin{equation}
P^{(i) \, ef}_{ab}P^{(j) \, cd}_{ef}= \delta^{ij} P^{(i) \, cd}_{ab}
\; ,
\end{equation}
and complete, and define a unique decomposition (modulo the $n(n+1)\over 2$
conformal Killing vectors in flat spacetime).

Because they are conserved tensors derived from invariant action
functionals, all the local tensors on the left hand side of the linear 
response eq.~(\ref{eq:linres}) must be expressible in terms of only
the scalar and tensor components of the metric fluctuations.
Indeed, by explicit computation in $n=4$ dimensions,
\begin{mathletters}
\begin{eqnarray}
\delta\, ^{(C)}H_{ab} &=& \sq^2 h^{(T)}_{ab} = \sq^2 
P_{ab}^{(T)\, cd}\,
h_{cd}
\; ,\\ 
\delta \,^{(1)}H_{ab} &=& 6 \sq^2 h^{(S)}_{ab} = 
6 \sq^2 P_{ab}^{(S)\, cd}\,
h_{cd}
\; ,
\\ 
\delta G_{ab} &=& \sq \left( - {1\over 2} h_{ab}^{(T)} +
h_{ab}^{(S)} \right) = -{1\over 2} \sq P_{ab}^{(T)\, cd}\, h_{cd} + \sq
P_{ab}^{(S)\, cd}\, h_{cd}
\; , 
\end{eqnarray}
\label{eq:wre}
\end{mathletters}

\noindent where $h_{ab} \equiv \delta g_{ab}$ is the metric perturbation
(variation).

The scalar and tensor projectors onto the space of gauge invariant
metric perturbations can be written in momentum space in the compact
forms,
\begin{mathletters}
\begin{eqnarray}
P^{(S) \, cd}_{ab}(k) &=& {1\over n-1} \theta_{ab} \theta^{cd}
\; ,
\\
P^{(T) \, cd}_{ab}(k) &=& {1\over 2} \left( \theta_a^c \theta_b^d  
+  \theta_a^d \theta_b^c \right) -{1\over n-1} \theta_{ab} \theta^{cd}
\; ,
\end{eqnarray}
\label{eq:projs}
\end{mathletters}

\noindent where we have introduced the tensor $\theta_{ab}$,
\begin{equation}
\theta_{ab} \equiv \eta_{ab} - {k_a k_b\over k^2} 
\; ,
\end{equation}
which obeys $k^a \theta_{ab} = k^b \theta_{ab} = 0$. Therefore,
the scalar and tensor projectors are also transverse:
\begin{equation}
k^b P^{(S) \, cd}_{ab}(k) = k^b P^{(T) \, cd}_{ab}(k)= 0
\; .
\end{equation}
Because of this property the correlation function of two conserved 
energy-momentum tensors in momentum space
\begin{equation}
\Pi_{ab}^{> \ cd} (k) = i\int\,{\d^nx\over (2\pi)^n} e^{ik\cdot(x-x')}
\langle T_{ab}(x) T^{cd}(x')\rangle
\; ,
\end{equation}
may be expanded in terms of the gauge invariant scalar and
tensor projectors only. By conservation of $T_{ab}$ this correlator
must be transverse, with zero projection onto the vector subspace,
{\it i.e.,}~\cite{Boul-Des}
\begin{equation}
\Pi_{ab}^{> \ cd}(k) = P^{(S) \, cd}_{ab}(k) \Pi^{(S)} (k) +
P^{(T) \ cd}_{ab} \Pi^{(T)} (k)
\label{func-s-t}
\; ,
\end{equation}
in terms of two scalar functions of momentum $k$. 

The most convenient way of expressing the retarded correlation function
in Fourier space is first to introduce the spectral function
representation~\cite{kal-leh} in the spin-2 (T) and spin-0 (S)
sectors, in terms of the Euclidean four-momentum $(k_4, \vec k)$
\begin{equation}
\Pi^{(i)}_{_E} (ik_4, \vec k) = 
\int_{-\infty}^{\infty} {\d\omega\over 2\pi}
{\sigma^{(i)}(\omega, \vec k) \over \omega - ik_4}
\; . 
\end{equation}
The retarded correlator is given then by the analytic continuation,
$ik_4 \rightarrow k^0 + i\epsilon$, {\it i.e.,}
\begin{equation}
\Pi^{(i)\, ({\rm ret})}(k^0, \vec k) = \Pi^{(i)}_{_E} (ik_4= k^0 + i\epsilon,
\vec k) = \int_{-\infty}^{\infty} {\d\omega\over 2\pi}
{\sigma^{(i)}(\omega, \vec k) \over \omega - k^0 - i\epsilon}
\; ,
\end{equation}
which corresponds to eq.~(\ref{eq:ac}) of the text.
Moreover, since $\sigma^{(i)} (\omega , \vec k)$ is an odd function of
$\omega$ (by the reality of $\Pi_{_E}^{(i)}$), which otherwise depends 
only on the Lorentz invariant combination $s \equiv \omega^2 - \vec k \cdot
\vec k$, we can define Lorentz invariant positive spectral functions
$\rho^{(i)} (s)$ by
\begin{equation}
\sigma^{(i)} (\omega, \vec k) = 2\pi\, {\rm sgn}(\omega)\, \rho^{(i)} (s)
\; ,
\label{eq:sigrho}
\end{equation}
and obtain the dispersion formula,
\begin{equation}
\Pi_{ab}^{({\rm ret}) \, cd}(k) = P^{(S) \, cd}_{ab}(k) \int_0^{\infty} 
{\d s \ \rho^{(S)}(s) \over s + k^2 - i \epsilon \; {\rm sgn} (k^0)} 
+  P^{(T) \,
cd}_{ab}(k) \int_0^{\infty} {\d s \ \rho^{(T)}(s) \over s + k^2 - i
\epsilon \; {\rm sgn} (k^0)}
\; , 
\label{eq:CLrep}
\end{equation}
which follows by substituting~(\ref{eq:sigrho}) into~(\ref{eq:ac}),
dividing the integration range over $\omega$ into positive and
negative $\omega$, renaming the integration variable, and using $\d s
= 2 \omega \d \omega$. Thus, the spectral functions $\rho^{(i)}(s)$ of
the two independent scalar functions given in eq.~(\ref{func-s-t}) can
be obtained by computing the simple correlator in Euclidean momentum
and evaluating the imaginary part, after performing the specific
analytic continuation~(\ref{eq:ac}), namely
\begin{equation}
{\rm Im}\left[\Pi^{(i)}_{_E}(ik_4 = k^0 + i\epsilon)\right]
 = \pi \, {\rm sgn}(k^0)\, 
\rho^{(i)}(s=-k^2)
\; ,
\label{eq:defrho}
\end{equation}
which is also obtained by continuing the Euclidean
\begin{equation}
(k_4)^2 + \vec k \cdot \vec k \rightarrow k^2 -i \epsilon \; {\rm sgn} (k^0)
\; ,
\end{equation}
to the Lorentzian $k^2=\eta_{ab}k^ak^b=-(k^0)^2 + \vec k \cdot \vec
k$. The usefulness of this representation is that the imaginary part
of the correlator is given by spectral functions which have
simple positivity properties and which are completely free of ultraviolet 
divergences. These appear only when the real part of the correlator is
constructed by the fully covariant dispersion integrals over $s$
in~(\ref{eq:CLrep}), and may be handled by standard methods that
make clear their relation to covariant local countertems in the
effective action. The covariant renormalization of these dispersion integrals
by explicit subtractions in flat space is described in Section~\ref{sec:4}.

\section{Gravitational Vacuum Polarization Tensor in Flat Space}

The classical energy-momentum tensor for a scalar field in $n$
dimensional Minkowski spacetime is given by
\begin{equation}
T_{ab}\big\vert_{\rm flat}= 
(1-2\xi)\nabla_a \Phi \nabla_b \Phi
+\left( 2\xi - \frac{1}{2} \right) \eta_{ab}\nabla_c \Phi \nabla^c \Phi
- 2 \xi \Phi \nabla_a  \nabla_b \Phi
+ 2 \xi \eta_{ab} \Phi \nabla_c  \nabla^c \Phi
-\frac{1}{2} \eta_{ab} m^2 \Phi^2
\; ,
\end{equation}
which can be rewritten as
\begin{equation}
T_{ab}\big\vert_{\rm flat}= 
\left[
\frac{1}{2} \partial_a \partial_b - \frac{1}{4}\eta_{ab} \sq 
- \frac{1}{2} m^2 \eta_{ab}
+ \xi \left(\eta_{ab} \sq - \partial_a \partial_b\right)
\right] \Phi^2
+ (- \delta_{a}^{c} \delta_{b}^{d}
+\frac{1}{2} \eta_{ab}\eta^{cd})
\Phi (\partial_c \partial_d \Phi)
\; .
\end{equation}
When this is substituted into the Fourier transform of the
energy-momentum tensor two-point connected correlation function in
Euclidean space and the two possible Wick contractions of $\langle
\Phi^2 (x) \Phi^2(x')\rangle_{\rm con}$ are taken into account, we
obtain
\begin{eqnarray}
\Pi_{ab}^{\ \ cd}\big\vert_E(k)&=& 
\int \d^n x \; \langle T_{ab}(x) T^{cd}(x') \rangle\big\vert_E 
e^{ik \cdot (x-x')} =
2 {\cal D}^{(1)}_{ab}(k) \; {\cal D}^{(1)cd}(k)\; H(k)
\nonumber
\\
&+&
2 {\cal D}^{(1)}_{ab}(k) \; {\cal D}^{(2)cd}_{c'd'} \; I^{c'd'}(k)
+
2 {\cal D}^{(2)a'b'}_{ab} \; {\cal D}^{(1)cd}(k) \; I_{a'b'}(k)
\nonumber
\\
&+&
2 {\cal D}^{(2)a'b'}_{ab} \; {\cal D}^{(2)cd}_{c'd'} \; 
J_{a'b'}^{\ \ c'd'}(k) +
2{\cal D}^{(2)a'b'}_{ab} \; {\cal D}^{(2)cd}_{c'd'} \; 
K_{a'b'}^{\ \ c'd'}(k)\; .
\label{pol-tensor}
\end{eqnarray}

\noindent Here $k=(k_4,\vec k)$ the Euclidean momentum, and we have
introduced the following tensors,
\begin{eqnarray}
{\cal D}^{(1)}_{ab}(k)&=& 
\xi(k_a k_b-\eta_{ab} k^2)
+
\frac{1}{4} (\eta_{ab} k^2 - 2 k_a k_b)
+ \frac{1}{2} m^2 \eta_{ab}
\; ,
\\
{\cal D}^{(2)cd}_{ab}&=&
- \delta_{a}^{c} \delta_{b}^{d}
+\frac{1}{2} \eta_{ab}\eta^{cd}
\; ,
\end{eqnarray}
and the notation for the following integrals,
\begin{eqnarray}
H(k)&=&
\int \frac{\d^n p}{(2\pi)^n}\frac{1}{p^2+m^2}\frac{1}{(p+k)^2+m^2}
\; ,
\\
I^{cd}(k)&=&
\int \frac{\d^n p}{(2\pi)^n}\frac{p^{c}p^{d}}{p^2+m^2}\frac{1}{(p+k)^2+m^2}
\; ,
\\
J_{ab}^{\ \ cd}(k)&=&
\int \frac{\d^n p}{(2\pi)^n}\frac{p_{a} p_{b} p^{c}p^{d}}
{p^2+m^2}\frac{1}{(p+k)^2+m^2}
\; ,
\\
K_{ab}^{\ \ cd}(k)&=&
\int \frac{\d^n p}{(2\pi)^n}\frac{p_{a} p_{b}}
{p^2+m^2}\frac{(p^{c}+k^{c})(p^{d}+k^{d})}{(p+k)^2+m^2}
\; .
\end{eqnarray}
We regularize all integrals by means of dimensional regularization in
$n$ dimensions~\cite{collins,dew-smith}. By introducing a Feynman
parameter $x$, the previous integrals are evaluated to
yield~\cite{dew-smith}
\begin{eqnarray}
H(k)&=&
\frac{\pi^{n/2}}{(2\pi)^n}\Gamma\left(2-\frac{n}{2} \right)
\int_0^1 \d x\ [m^2 + k^2 x(1-x)]^{\frac{n}{2}-2}
\; ,
\\
I^{cd}(k)&=&
\frac{\pi^{n/2}}{(2\pi)^n}\Gamma\left(2-\frac{n}{2} \right)
\int_0^1 \d x\ [m^2 + k^2 x(1-x)]^{\frac{n}{2}-2}
\left\{
\frac{\eta^{cd}}{2-n}[m^2 + k^2 x(1-x)]+x^2 k^c k^d
\right\}
\; ,
\\
J_{ab}^{\ \ cd}(k)&=&
\frac{\pi^{n/2}}{(2\pi)^n}\Gamma\left(2-\frac{n}{2} \right)
\int_0^1 \d x\ [m^2 + k^2 x(1-x)]^{\frac{n}{2}-2}
\nonumber
\\
&\times&
\left\{
\frac{1}{n(n-2)} (\tau^{(1)cd}_{ab}+\tau^{(2)cd}_{ab})
[m^2 + k^2 x(1-x)]^2
-
\frac{x^2}{(n-2)} (\tau^{(3)cd}_{ab}+\tau^{(4)cd}_{ab})
[m^2 + k^2 x(1-x)]
\right.
\nonumber
\\
&&
\qquad + \left.
x^4\tau^{(5)cd}_{ab}
\right\}
\; ,
\\
K_{ab}^{\ \ cd}(k)&=&
\frac{\pi^{n/2}}{(2\pi)^n}\Gamma\left(2-\frac{n}{2} \right)
\int_0^1 \d x\ [m^2 + k^2 x(1-x)]^{\frac{n}{2}-2}
\nonumber
\\
&\times&
\left\{
\frac{1}{n(n-2)} (\tau^{(1)cd}_{ab}+\tau^{(2)cd}_{ab})
[m^2 + k^2 x(1-x)]^2
- \frac{x^2}{(n-2)} \tau^{(3)cd}_{ab}
[m^2 + k^2 x(1-x)]
\right.
\nonumber
\\
&&
\qquad +\left.
\frac{x(1-x)}{(n-2)} \tau^{(4)cd}_{ab}
[m^2 + k^2 x(1-x)]
+x^2(1-x)^2\tau^{(5)cd}_{ab}
\right\}
\; . 
\end{eqnarray}
These expressions are given in terms of the following five basis tensors,
\begin{eqnarray}
\tau^{(1)cd}_{ab}(k)&=& \eta_{ab}\eta^{cd}
\; ,
\\
\tau^{(2)cd}_{ab}(k)&=& \delta_a^c \delta_b^d+\delta_a^d \delta_b^c
\; ,
\\
\tau^{(3)cd}_{ab}(k)&=&\eta_{ab} k^c k^d+\eta^{cd} k_a k_b
\; ,
\\
\tau^{(4)cd}_{ab}(k)&=&
\delta_a^c k_b k^d+\delta_a^d k_b k^c +\delta_b^c k_a k^d + \delta_b^d k_a k^c
\; ,
\\
\tau^{(5)cd}_{ab}(k)&=&k_ak_bk^ck^d
\; .
\end{eqnarray}
In order to obtain the Euclidean polarization tensor we need to
compute the corresponding tensor products specified in
equation~(\ref{pol-tensor}).  Once we do this we can write the
polarization operator (in $n$ dimensions and for $\xi=0$) as
\begin{eqnarray}
\Pi_{ab}^{\ \ cd}\big\vert_E (k,\xi=0)&=&
\sum_{j=1}^{5} F_j(k) \tau^{(j)cd}_{ab}(k)
\; ,
\end{eqnarray}
where
\begin{eqnarray}
F_1(k)
&=&
\frac{\pi^{n/2}}{(2\pi)^n}\Gamma\left(2-\frac{n}{2} \right)
\int_0^1 \d x \; [m^2 + k^2 x(1-x)]^{\frac{n}{2}-2}
\nonumber
\\
&\times&
\left[
\frac{n^2-2n-4}{2n(n-2)}[m^2 + k^2 x(1-x)]^2 
+\left(\frac{xk^2}{n-2}(1-nx) + \frac{k^2}{2}-m^2 \right)
[m^2 + k^2 x(1-x)]
\right.
\nonumber
\\
&+&
\left.
\frac{k^4x^2}{4}\left((1-x)^2+x^2\right)
+k^2x^2 \left(-\frac{k^2}{2}+m^2\right)+\frac{1}{2}
\left(-\frac{k^2}{2}+m^2\right)^2
\right]
\; ,
\\
F_2(k)&=&
\frac{\pi^{n/2}}{(2\pi)^n}\Gamma\left(2-\frac{n}{2} \right)
\int_0^1 \d x \; [m^2 + k^2 x(1-x)]^{\frac{n}{2}-2}
\frac{2}{n(n-2)}[m^2 + k^2 x(1-x)]^{2}
\; ,
\\
F_3(k)&=&
\frac{\pi^{n/2}}{(2\pi)^n}\Gamma\left(2-\frac{n}{2} \right)
\int_0^1 \d x \; [m^2 + k^2 x(1-x)]^{\frac{n}{2}-2}
\nonumber
\\
&\times&
\left\{
\left[
\frac{-2x+(n+2)x^2}{(n-2)} -\frac{1}{2}\right][m^2 + k^2 x(1-x)]
-\frac{k^2}{4}+\frac{m^2}{2}
-\frac{k^2}{2}x^2\left((1-x)^2+x^2-2\right) -x^2 m^2
\right\}
\; ,
\\
F_4(k)&=&
\frac{\pi^{n/2}}{(2\pi)^n}\Gamma\left(2-\frac{n}{2} \right)
\int_0^1 \d x \; [m^2 + k^2 x(1-x)]^{\frac{n}{2}-2}
\frac{x(1-2x)}{(n-2)}[m^2 + k^2 x(1-x)]
\; ,
\\
F_5(k)&=&
\frac{\pi^{n/2}}{(2\pi)^n}\Gamma\left(2-\frac{n}{2} \right)
\int_0^1 \d x \; [m^2 + k^2 x(1-x)]^{\frac{n}{2}-2}
\left[
\frac{1}{2}+x^2((1-x)^2+x^2-2)
\right]
\; .
\end{eqnarray}
The $\xi$ dependent part of the polarization operator can be written as
\begin{eqnarray}
&&\Pi_{ab}^{\ \ cd}\big\vert_E (k,\xi)= 
\xi
\frac{\pi^{n/2}}{(2\pi)^n}\Gamma\left(2-\frac{n}{2} \right)
\int_0^1 \d x \; [m^2 + k^2 x(1-x)]^{\frac{n}{2}-2} 
\nonumber
\\
&\times&
\left[
k^2 (\theta_{ab}\eta^{cd}
+\theta^{cd}\eta_{ab})
\left(
-[m^2 + k^2 x(1-x)] - \frac{k^2}{2}+m^2 +k^2 x^2 \right) 
+(\theta_{ab}k^c k^d +k_ak_b \theta^{cd})
k^2 (1-2x^2)
\right]
\nonumber
\\
&+&
2\xi^2 k^4 \theta_{ab}\theta^{cd}
\frac{\pi^{n/2}}{(2\pi)^n}\Gamma\left(2-\frac{n}{2} \right)
\int_0^1 \d x \; [m^2 + k^2 x(1-x)]^{\frac{n}{2}-2} 
\; . 
\end{eqnarray} 
In order to obtain the retarded polarization operator we next 
analytically continue the Euclidean momentum to a Lorentzian momentum.
This analytic continuation is defined by 
\begin{equation}
k_4^2+\vec k^2 \rightarrow
\eta_{ab}k^ak^b -i \, \epsilon \, {\rm sgn} (k^0) \;, 
\end{equation}
with $\epsilon \rightarrow 0^+$ (See Appendix A). Notice that the continuation
depends on the sign of $k^0$. We also take the limit $n\rightarrow 4$
and write $n =4-\delta$, with $\delta \rightarrow 0^+$. The real part
of the polarization tensor has a pole at $n=4$, but its imaginary part
comes only from the logarithmic branch cut of the function, $[m^2 +
k^2 x(1-x)]^{-\delta/2}$ expanded around $\delta =0$,
\begin{eqnarray}
{\rm Im} \left(
[m^2+k^2 x (1-x)]^{-\delta/2}
\right) = \frac{\pi\delta}{2}\;
{\rm sgn}(k^0) \;
 \theta[-m^2-\eta_{ab}k^ak^bx(1-x)] + {\cal O}(\delta^2)
\; ,
\end{eqnarray}
so that the pole in $\Gamma (\delta/2) \rightarrow 2/\delta$ at
$\delta =0$ is canceled.  Thus the imaginary part is finite in the
limit $n \rightarrow 4$. Because $x(1-x) \le 1/4$ in the interval
$[0,1]$, the step function condition is satisfied only if
$\frac{s}{4}-m^2>0$, where $s=-k^2 = \eta_{ab}k^ak^b$, and $x \in
[x_{-},x_{+}]$, with $x_{\pm}$ the real roots of the quadratic
polynomial $m^2-x(1-x)s=0$. Specifically,
\begin{eqnarray} x_{\pm}= \frac{1}{2}(1\pm r) \; ,
\; \; \; {\rm with} \; \; \; 
r\equiv \sqrt{1-\frac{4m^2}{s}}
\; .
\end{eqnarray}
Therefore,
\begin{eqnarray}
{\rm Im} \left(
\lim_{\delta \rightarrow 0^+} 
\int_0^1 \d x \left[  \Gamma\left(\frac{\delta}{2}\right)
[m^2+k^2 x (1-x)]^{-\delta/2} \right]\dots
\right)
&=&
\int_{x_{-}}^{x_{+}} \d x \; \pi \; {\rm sgn}(k^0) \; \theta(s-4m^2)\dots
\; ,
\end{eqnarray}
where the ellipsis denotes any function of $x$ to be integrated.

Thus, the $x$ integrals all become simple powers of $x$, and for the
case $n=4$, we obtain
\begin{eqnarray}
{\rm Im} [F_1(s)]
 &=& 
{\rm sgn}(k^0) \,\frac{\theta(s-4m^2)}{16\pi} \;
\frac{r}{15}\left(2m^4 + 4m^2 s +\frac{3}{4}s^2  \right)
\; ,
\\
{\rm Im} [F_2(s)] &=&
{\rm sgn}(k^0) \, \frac{\theta(s-4m^2)}{16\pi} \;
\frac{r}{15}\left(2m^4 -m^2 s +\frac{1}{8}s^2  \right)
\; ,
\\
{\rm Im} [F_3(s)] &=&
{\rm sgn}(k^0) \, \frac{\theta(s-4m^2)}{16\pi} \;
\frac{r}{15s}\left(2m^4 +4m^2 s +\frac{3}{4}s^2  \right)
\; ,
\\
{\rm Im} [F_4(s)]&=&
{\rm sgn}(k^0) \, \frac{\theta(s-4m^2)}{16\pi} \;
\frac{r}{15s}\left(2m^4 -m^2 s +\frac{1}{8}s^2  \right)
\; ,
\\
{\rm Im} [F_5(s)]&=&{\rm sgn}(k^0) \, \frac{\theta(s-4m^2)}{16\pi} \;
\frac{r}{15s^2}\left(6m^4 +2m^2 s +s^2  \right)
\; .
\end{eqnarray}
These functions are not linearly independent, as the
covariant conservation of the energy-momentum tensor implies that
\begin{mathletters}
\begin{eqnarray}
F_1(s) - s F_3(s) &=& 0
\; ,
\\
F_2(s) - s F_4(s) &=& 0
\; ,
\\
F_3(s) +2 F_4(s)- s F_5(s) &=& 0
\; .
\end{eqnarray}
\end{mathletters}

\noindent In fact the combinations, $2F_2$ and $3F_1 + 2F_2$, yield the tensor
and scalar spectral functions of eq.~(\ref{eq:CLrep}), respectively. 

Likewise, after computing the $x$ integrals, the $\xi$ dependent polarization
tensor becomes 
\begin{eqnarray}
{\rm Im} \left[\Pi_{ab}^{\ \ cd}(s,\xi)\right]&=&
{\rm sgn} (k^0) \,\frac{\theta(s-4m^2)}{24\pi}\,
\sqrt{1-\frac{4m^2}{s}}\,
\theta_{ab}\theta^{cd}\, \left[
-\xi s (s+2m^2) + 3\xi^2 s^2 
\right]
\; ,
\end{eqnarray}
which is explicitly transverse and proportional to the scalar
projector $P^{(S)\,cd}_{ab}$. Combining this $\xi$ dependent
contribution with the previous $\xi$ independent part, and
recalling~(\ref{eq:projs}) and~(\ref{eq:defrho}), we may now identify
the two independent tensor and scalar spectral functions,
\begin{mathletters} 
\begin{eqnarray}
\rho^{(\rm T)}(s)&=&
\frac{\theta(s-4m^2)}{60 \pi^2}\,  \sqrt{1-\frac{4m^2}{s}}\,
\left(\frac{s}{4}-m^2\right)^2
\ge 0
\; ,
\label{eq:rho-T}
\\
\rho^{\rm (S)}(s)&=&
\frac{\theta(s-4m^2)} {24 \pi^2} \, \sqrt{1-\frac{4m^2}{s}}
\,\left[ 
m^2 +\frac{(1-6 \xi)s}{2}
\right]^2
\ge 0
\; .
\label{eq:rho-S}
\end{eqnarray}
\end{mathletters}

\noindent Both spectral functions are positive, as they must be,
and agree with results (for $\xi =0$) reported in~\cite{jaekel-reynaud},
and (for arbitrary $m$ and $\xi$) reported in~\cite{martin-verdaguer-3}

In the case that the curvature coupling takes its conformal value,
$\xi=1/6$, the scalar spectral function does not have terms
proportional to $s^2$ or to $m^2s$, and becomes
\begin{eqnarray}
\rho^{(\rm S)}\big\vert_{\xi=1/6}(s)&=&
\theta(s-4m^2) \frac{m^4}{24 \pi^2}  \sqrt{1-\frac{4m^2}{s}}
\label{eq:rho-S-con}
\; ,
\end{eqnarray}
which agrees with~\cite{mazmot}, after account is taken of a
relative factor of $2m^4/3$ in the definition of the spectral 
function $\rho^{(S)}(s)$ here, relative to $\rho(s)$ of that work.

Finally the integrals appearing in the K\"all\'en-Lehmann
representations~(\ref{eq:CLrep}) are all of the form,
\begin{equation}
I_{n, l} \equiv k^2 \int_{4m^2}^{\infty} {\d s \over s^{l+1} (s+ k^2)}
\left( 1 - {4m^2\over s}\right)^{n + {1\over 2}}
\end{equation}
for $k^2 > 0$ and $n$ and $l$ integers. By making the change of variables
$s = 4m^2/(1-u^2)$, all integrals of this kind may be reduced to linear
combinations of
\begin{equation}
I_n(z) \equiv I_{n, l=0} = 2 \int_0^1\, \d u {u^{2n+2}\over z^2 - u^2}
\; ,
\end{equation}
where
\begin{equation}
z \equiv \sqrt{1 + {4m^2\over k^2}}\,. 
\; ,
\end{equation}
For $l=0$ the $I_n(z)$ functions obey the recursion formula,
\begin{equation}
I_n(z) = -{2\over 2n +1} + z^2 I_{n-1}(z)
\; ,
\end{equation}
with
\begin{eqnarray}
I_0(z) &=& -2 + z \log 
\left({z+1\over z-1}\right)\qquad {\rm for} \qquad z>1
\nonumber\\
&\equiv& -2 + f\left({k^2\over m^2}\right)\qquad {\rm for} \qquad k^2>0
\; .
\end{eqnarray}
Using these relations, the response function for the tensor
fluctuations can be written as
\begin{eqnarray}
F^{(T)}\left({k^2\over m^2}\right) &\equiv & k^2 \int_{4m^2}^{\infty} {\d s
\over s^3 (s+ k^2)}\ \rho^{(T)}(s)  \nonumber\\
&=& {1\over 960\pi^2} \left[ -{2\over 5} - {2 \over 3}z^2 + z^4 I_0(z)\right]
\nonumber\\
&=& {1\over 960\pi^2} \left[ -{46\over 15} - {56\over 3}{m^2\over k^2}
-{32 m^4\over (k^2)^2} + \left( 1 + {4m^2\over k^2}\right)^2
f\left({k^2\over m^2}\right)\right]
\; ,
\label{eq:tenF}
\end{eqnarray}
while the corresponding response function for the scalar fluctuations is
\begin{eqnarray}
F^{(S)} \left({k^2\over m^2}\right) &\equiv & k^2 \int_{4m^2}^{\infty} {\d s
\over s^3 (s+ k^2)}\ \rho^{(S)}(s) \nonumber\\
&=& {1\over 96\pi^2} \left\{ {1\over 15} + {2 \over 3}(1-6\xi) -{2\over 3}
{m^2\over k^2} + \left[ (1- 6\xi) - {2m^2\over k^2}\right]^2
\left[-2 + f\left({k^2\over m^2}\right)\right] \right\}
\; .
\label{eq:scaF}
\end{eqnarray}
As $k^2 \rightarrow 0$, $z\rightarrow \infty$, and the function $I_0$
(or $f$) is analytic at $z^{-1}=0$. However, as $k^2$ changes sign,
$z^{-1}$ becomes pure imaginary and
\begin{equation}
f\left({k^2\over m^2}\right)= 2\left({4m^2 \over s}-1\right)^{1\over
2}\,\tan^{-1} \left[\left({4m^2 \over s}-1\right)^{-{1\over 2}}\right]
\qquad {\rm for} \qquad 0 < s = -k^2 \le 4m^2
\; ,
\end{equation}
which remains real in this range. Finally when $s = -k^2 > 4m^2$, $f$
develops an imaginary part,
\begin{equation}
f\left({k^2\over m^2}\right)= z \log \left({1+z\over 1-z}\right)
-i \pi z \ {\rm sgn}(k^0)\qquad {\rm for}
\qquad z = \sqrt {1 - {4m^2\over s}}\; , \ s= -k^2 \ge 4m^2\; ,
\ 0\le z<1
\; .
\end{equation}


\begin{references}

\bibitem{hu} 
B.-L. Hu, Physica A {\bf 158}, 399 (1989).

\bibitem{donoghue} 
J. F. Donoghue, Phys. Rev. D {\bf 50}, 3874 (1994).

\bibitem{birrell-davies}
N. D. Birrell and P. C. W. Davies,
{\it Quantum Fields in Curved Space},
Cambridge University Press (Cambridge), 1982,
and references therein.

\bibitem{FW}
See {\it e.g.,} A. L. Fetter and J. D. Walecka, {\it Quantum Theory of
Many-Particle Systems}, McGraw-Hill (New York), 1971.

\bibitem{leb}
See {\it e.g.,} J. I. Kapusta, {\it Finite Temperature Field Theory},
Cambridge University Press (Cambridge), 1989; \hfill\break
M. Le Bellac, {\it Thermal Field Theory}, Cambridge University Press
(Cambridge), 1996. 

\bibitem{mot-fd}
E. Mottola, Phys. Rev. D {\bf 33}, 2136 (1986).

\bibitem{dewitt}
B. S. DeWitt,
{\it  in Les Houches 1985, Proceedings, Architecture Fundamental Interactions 
at Short Distances, Vol. 2, 1023-1057.} 

\bibitem{dw-mp}
B. S. DeWitt and C. Molina-Par\'\i s,
Mod. Phys. Lett. A {\bf 13}, 2475 (1998).

\bibitem{brandenberger}
See however V. Mukhanov, H. Feldman and R. Brandenberger, 
Phys. Rep. {\bf 215}, 203 (1992) for developments in a cosmological context.

\bibitem{tomboulis}
E. Tomboulis, Phys. Lett. B {\bf 70}, 361 (1977).

\bibitem{mtw}
Our sign and curvature conventions are those of C. W. Misner, K. S. Thorne,
and J. A. Wheeler, {\it Gravitation} (Freeman, San Francisco, 1973).

\bibitem{ashtekar}  
A. Ashtekar,  
Phys. Rev. Lett. {\bf 77}, 4864 (1996). 
 
\bibitem{beetle}  
C. Beetle, Adv. Theor. Math. Phys. {\bf 2}, 471 (1998). 

\bibitem{ford}
L. H. Ford, Ann. Phys. (N.Y.) {\bf 144}, 238 (1982). 

\bibitem{kuo-ford} 
C.-I. Kuo and L. H. Ford, 
Phys. Rev. D {\bf 47}, 4510 (1993).

\bibitem{wu-ford-1}
C.-H. Wu and L. H. Ford, 
Phys. Rev. D {\bf 60}, 104013 (1999).

\bibitem{wu-ford-2} 
C.-H. Wu and L. H. Ford, Phys. Rev. D {\bf 64}, 045010 (2001);
e-print gr-qc/0102063.

\bibitem{wu-ford-3} 
L. H. Ford and C.-H. Wu, e-print gr-qc/0102063.

\bibitem{phillips-hu-1} 
N. G. Phillips and B.-L. Hu, 
Phys. Rev. D {\bf 55}, 6123 (1997).

\bibitem{phillips-hu-2} 
B.-L. Hu and N. G. Phillips, 
Int. J. Theor. Phys. {\bf 39}, 1817 (2000);\hfill\break
N. G. Phillips and B. L. Hu, Phys. Rev. D {\bf 62}, 084017 (2000).

\bibitem{calzetta-hu}
E. Calzetta and B.-L. Hu, 
Phys. Rev. D {\bf 49}, 6636 (1994).

\bibitem{hu-matacz}
B.-L. Hu and A. Matacz, 
Phys. Rev. D {\bf 51}, 1577 (1995).

\bibitem{hu-sinha}
B.-L. Hu and S. Sinha, 
Phys. Rev. D {\bf 51}, 1587 (1995).

\bibitem{martin-verdaguer-1}
R. Mart\'{\i}n and E. Verdaguer,
Phys. Lett. B {\bf 465}, 113 (1999).

\bibitem{martin-verdaguer-2}
R. Mart\'{\i}n and E. Verdaguer,
Phys. Rev. D {\bf 61}, 084008 (1999).

\bibitem{campos-hu-1}
A. Campos and B.-L. Hu, 
Phys. Rev. D {\bf 58}, 125021 (1998).

\bibitem{carlitz-willey} 
R. D. Carlitz and R. S. Willey, 
Phys. Rev. D {\bf 36}, 2327 (1987).

\bibitem{jaekel-reynaud}
M.-T. Jaekel and S. Reynaud,
Annalen Phys. {\bf 4}, 68 (1995).

\bibitem{martin-verdaguer-3} 
R. Mart\'{\i}n and E. Verdaguer,
Phys. Rev. D {\bf 61}, 124024 (2000). 

\bibitem{roura-verdaguer} 
A. Roura and E. Verdaguer, 
Int. J. Theor. Phys. {\bf 38}, 3123 (1999).

\bibitem{capper-duff-halpern}
D. M. Capper, M. J. Duff, and L. Halpern,
Phys. Rev. D {\bf 10}, 461 (1974).

\bibitem{horowitz}
G. T. Horowitz, 
Phys. Rev. D {\bf 21}, 1445 (1980).

\bibitem{jordan-1}
R. D. Jordan,
Phys. Rev. D {\bf 36}, 3593 (1987).

\bibitem{hartle-hu}
J. B. Hartle and B.-L. Hu,
Phys. Rev. D {\bf 20}, 1772 (1979);
{\it ibid.} 
{\bf 21}, 2756 (1979);
J. B. Hartle,
Phys. Rev. D {\bf 22}, 2091 (1980).

\bibitem{horowitz-wald}
G. T. Horowitz and R. M. Wald,
Phys. Rev. D {\bf21}, 1462 (1980).

\bibitem{campos-verdaguer-1}
A. Campos and E. Verdaguer,
Phys. Rev. D {\bf 53}, 1927 (1996).

\bibitem{campos-verdaguer-2}
A. Campos and E. Verdaguer,
Int. J. Theor. Phys. {\bf 36}, 2525 (1997).

\bibitem{campos-hu-2}
A. Campos and B.-L. Hu, 
Int. J. Theor. Phys. {\bf 38}, 1253 (1999).

\bibitem{calzetta-campos-verdaguer}
E. Calzetta, A. Campos, and E. Verdaguer,
Phys. Rev. D {\bf 56}, 2163 (1997).

\bibitem{calzetta-verdaguer}
E. Calzetta and E. Verdaguer,
Phys. Rev. D {\bf 59}, 083513 (1999).

\bibitem{lombardo-mazzitelli} 
F. C. Lombardo and F. D. Mazzitelli,
Phys. Rev. D {\bf 55}, 3889 (1997).

\bibitem{mar-sch}
P. C. Martin and J. Schwinger, Phys. Rev. {\bf 115}, 1342 (1959).

\bibitem{CTP}
J. Schwinger, J. Math. Phys. {\bf 2}, 407 (1961);  \hfill\break                
L. V. Keldysh, Zh. Eksp. Teor. Fiz. {\bf 47}, 1515 (1964)
[Sov. Phys. JETP {\bf 20}, 1018 (1965)];  \hfill\break
K.-C. Chou, Z.-B. Su, B.-L. Hao, and L. Yu,
Phys. Rep. {\bf 118}, 1 (1985).

\bibitem{photon}
See {\it e.g.,} C. Itzykson and J.-B. Zuber, {\it Quantum Field Theory},
McGraw-Hill (New York), 1980.

\bibitem{kal-leh}
G. K\"all\'en, {\it Quantum Electrodynamics}, Springer-Verlag
(Berlin), 1972.

\bibitem{suen}
W.-M. Suen,
Phys. Rev. Lett. {\bf 62}, 2217 (1989);
Phys. Rev. D {\bf 40}, 315 (1989).

\bibitem{simon} 
J. Z. Simon, Phys. Rev. D {\bf 41}, 3720 (1990); 
{\it ibid.} 
{\bf 43}, 3308 (1991).

\bibitem{parker-simon}
L. Parker and J. Z. Simon,
Phys. Rev. D {\bf 47}, 1339 (1993).

\bibitem{flanagan-wald}
E. E. Flanagan and R. M. Wald,
Phys. Rev. D {\bf 54}, 6233 (1996).

\bibitem{donoghue}
J. F. Donoghue, Phys. Rev. D {\bf 50}, 3874 (1994). 

\bibitem{CHKM}
F. Cooper, S. Habib, Y. Kluger, E. Mottola, J. P. Paz,
and P. R. Anderson, Phys. Rev. D {\bf 50}, 2848 (1994);\hfill\break
F. Cooper, S. Habib, Y. Kluger, and E. Mottola, Phys. Rev.
D {\bf 55}, 6471 (1997).

\bibitem{jordan-2}
R. D. Jordan, Phys. Rev. D {\bf 33}, 444 (1986).

\bibitem{har-sch}
J. B. Hartle and K. Schleich, in {\it Quantum Field Theory and Quantum
Statistics}, T. A. Batalin, C. J. Isham, and G. A. Vilkovisky, eds.,
Hilger (Bristol), 1988;\hfill\break
K. Schleich, Phys. Rev. D {\bf 36}, 2342 (1987)

\bibitem{mazmot}
P. O. Mazur and E. Mottola, Nucl. Phys. B {\bf 341}, 187 (1990).

\bibitem{rebhan}
H. Nachbagauer, A. K. Rebhan, and D. J. Schwarz,
Phys. Rev. D {\bf 53}, 5468 (1996).

\bibitem{mot-des}
E. Mottola, Phys. Rev. D {\bf 31}, 754 (1985);
{\it ibid.} {\bf 33}, 1616 (1986);\hfill\break
P. O. Mazur and E. Mottola, Nucl. Phys. B {\bf 278}, 694 (1986);\hfill\break
I. Antoniadis and E. Mottola, J. Math. Phys. {\bf 32}, 1037 (1991);
\hfill\break
E. Mottola, J. Math. Phys. {\bf 36}, 2470 (1995).

\bibitem{harhaw}
J. B. Hartle and S. W. Hawking, Phys. Rev. D {\bf 13}, 2188 (1976).

\bibitem{york} 
J. W. York, Jr., Phys. Rev. D {\bf 31}, 775 (1985).

\bibitem{hky} 
D. Hochberg, T. W. Kephart, and J.W. York, Jr., 
Phys. Rev. D {\bf 48}, 479 (1993).

\bibitem{ahwy} 
P. R. Anderson, W. A. Hiscock, J. Whitesell, and J. W. York Jr.,
Phys. Rev. D {\bf 50}, 6427 (1994).

\bibitem{unstable}
S. Hawking, Phys. Rev. D {\bf 13}, 191 (1976).

\bibitem{Boul-Des}
D. G. Boulware and S. Deser, Jour. Math. Phys. {\bf 8}, 1468 (1967);\hfill\break
L. S. Brown and A. Zee, Jour. Math. Phys. {\bf 24}, 1822
(1983).

\bibitem{collins} See {\it e.g.,} J. C. Collins, {\it
Renormalization}, Cambridge University Press (Cambridge), 1984.

\bibitem{dew-smith}
See {\it e.g.,} B. De Wit and J. Smith, {\it Field Theory in Particle Physics},
North-Holland (Amsterdam), 1986.


\end{references}
\end{document}